\documentclass[twocolumn]{aastex631}

\submitjournal{ApJ}

\usepackage{hyperref}
\usepackage{graphicx}

\received{October 11, 2023}
\accepted{January 04, 2024}  

\begin{document}

\title{Yet Another Sunshine Mystery:\\Unexpected Asymmetry in GeV Emission from the Solar Disk}


\correspondingauthor{Bruno Arsioli, Elena Orlando}
\email{bruno.arsioli@gmail.com, orland.ele@gmail.com}

\author[0000-0001-8166-6602]{Bruno Arsioli}
\affiliation{Institute of Astrophysics and Space Science, OAL, University of Lisbon, Tapada da Ajuda, 1349-018 Lisbon, Portugal}

\affiliation{Department of Physics, Faculty of Sciences, University of Lisbon, Edifício C8, Campo Grande, 1749-016 Lisbon, Portugal}

\affiliation{University of Trieste, Department of Physics Via Valerio 2, 34127 Trieste, Italy}

\affiliation{Istituto Nazionale di Fisica Nucleare - INFN Padriciano, 99, 34149 Trieste, Italy}

\author{Elena Orlando}

\affiliation{University of Trieste, Department of Physics Via Valerio 2, 34127 Trieste, Italy}

\affiliation{Istituto Nazionale di Fisica Nucleare - INFN Padriciano, 99, 34149 Trieste, Italy}

\affiliation{Stanford University, 94305 Stanford, California, USA}

\begin{abstract}
The Sun is one of the most luminous $\gamma$-ray sources in the sky and continues to challenge our understanding of its high-energy emission mechanisms. This study provides an in-depth investigation of the solar disk $\gamma$-ray emission, using data from the Fermi Large Area Telescope spanning 2008 August to 2022 January. We focus on $\gamma$-ray events with energies exceeding 5 GeV, originating from 0.5$^\circ$ angular aperture centered on the Sun, and implement stringent time cuts to minimize potential sample contaminants. We use a helioprojection method to resolve the $\gamma$-ray events relative to the solar rotation axes and combine statistical tests to investigate the distribution of events over the solar disk. We found that integrating observations over large time windows may overlook relevant asymmetrical features, which we reveal in this work through a refined time-dependent morphological analysis. We describe significant anisotropic trends and confirm compelling evidence of energy-dependent asymmetry in the solar disk $\gamma$-ray emission. Intriguingly, the asymmetric signature coincides with the Sun's polar field flip during the cycle 24 solar maximum, around 2014 June. Our findings suggest that the Sun's magnetic configuration plays a significant role in shaping the resulting $\gamma$-ray signature, highlighting a potential link between the observed anisotropies, solar cycle, and the solar magnetic fields. These insights pose substantial challenges to established emission models, prompting fresh perspectives on high-energy solar astrophysics.

\end{abstract}

\keywords{Quiet sun (1322) ; Solar gamma-ray emission (1497) ; Solar cycle (1487) ; Solar magnetic fields (1503) ; High energy astrophysics (739)}



\section{Introduction}

The Sun, a persistent and intense source of $\gamma$-ray emission, manifests a plethora of processes within its solar atmosphere. Efforts to model and quantify the solar $\gamma$-ray flux date back to \cite{Seckel-CR-Sun-1991}, a time when detecting the solar disk's $\gamma$-ray signature was thought to be within the capabilities of forthcoming detectors like the EGRET on board the Compton Gamma Ray Observatory \citep{CGRO-Mission}. Since then, one of the primary mechanisms believed to fuel this $\gamma$-ray emission is the interaction between high-energy cosmic rays (CRs) and the Sun's outer layers, as well as its complex magnetic field structure. Specifically, the collisions of high-energy CRs with the solar atmosphere produce fluxes of $\gamma$-rays, neutrinos, antiprotons, neutrons, and antineutrons. The intensities of these fluxes hinge on the models of CR motion in the inner solar system's magnetic fields \citep[typically portrayed by Parker's spiral structure;][]{Parker-1965} and the postulated magnetic configurations on the solar surface. Indeed, both modeling the solar $\gamma$-ray emissions and reconciling them with observations remain formidable challenges to this day. 

Investigating solar $\gamma$-ray emission presents unique challenges due to the Sun's transit against a background populated with point and extended sources, including the diffuse $\gamma$-ray emission associated with our Galaxy's disk. This transit introduces a non-isotropic and time-variability component to the solar background, adding complexity to the analysis. Despite these analytical hurdles, \cite{ElenaOrlando-Sun-EGRET-2008} first successfully detected the solar $\gamma$-ray signature using EGRET observations, revealing the components tied to the solar disk and solar halo. Upper limits with EGRET to the solar disk emission were previously obtained by \cite{Thompson}. Subsequently, with the launch of the Fermi Large Area Telescope (LAT) mission \citep{FermiLAT}, \cite{ElenaOrlando-QuiescentSun-2009} and \cite{FermiLAT-QuietSun-2011} were able to probe the solar $\gamma$-ray emission with improved resolution and sensitivity, providing a better identification of the disk and halo spectral components. 

Building on these foundational studies, \cite{TimLinden-2018} made significant advancements in our understanding. Their analysis used helioprojective coordinates to investigate the localization of $\gamma$-ray events from the solar disk, providing an unprecedented view of the emission morphology. Their work include the identification of different morphological components of the $\gamma$-ray emission (with variation over the solar cycle, and different spectral features) and the detection of solar disk $\gamma$-ray flux at energies above 100 GeV during the solar minimum (of cycle 24). They employed a Kolmogorov-Smirnov (KS) test to compare event distributions against a model of constant surface brightness, and for the first time unveiled deviations from the expected isotropic disk emission. Furthermore, \cite{TimLinden-2022} delivered a $\gamma$-ray analysis of the solar disk over an entire 11 yr solar cycle and presented a detailed spectrum between 100 MeV and 100 GeV. Their work unveiled a strong anticorrelation between solar activity and $\gamma$-ray emission, highlighting the role of solar atmospheric magnetic fields over cosmic-ray modulation, which influences both the $\gamma$-ray flux and morphology.

The disk and halo components of solar $\gamma$-ray emission manifest distinct spatial and spectral characteristics. The compact disk emission corresponds to the solar disk angular size (of approximately $0.26^\circ$ radius), while the halo emission extends to large angular distances around the Sun, generating a significant diffuse foreground in the $\gamma$-ray sky \citep{ElenaOrlando-StellarICS-2020}. The solar halo component is primarily attributed to inverse-Compton (IC) scattering of the solar radiation field by relativistic cosmic-ray electrons. This IC component was first modeled by \cite{ElenaOrlando-Gamma-StarHalos-2006,ElenaOrlando-Sun-EGRET-2008} and predicted to yield a flux (F$_{\geq100 MeV}$) of 2.18$\times$10$^{-7}$cm$^{-2}$s$^{-1}$ within 10$^\circ$ of the Sun's center. Both EGRET and Fermi-LAT observations have confirmed this solar halo flux, aligning with the predicted estimates \citep{FermiLAT-QuietSun-2011}. Other mechanisms, such as particle acceleration in solar flares and coronal mass ejections \citep{Chupp-1973,Kanbash-1993,FermiLAT-QuietSun-2011,SunFlare-Share2017,FermiLAT-SolarFlareCat-2021}, are also known to induce intense $\gamma$-ray emissions, which can last from several minutes to hours, but with steep spectra.

The emission from the solar disk is anticipated to be mainly driven by hadronic interactions of CRs with the solar surface. \cite{Seckel-CR-Sun-1991} argued that these CRs are predominantly protons and discussed the influence of the solar magnetic field over them. Cosmic-ray protons with energy exceeding a certain threshold ($E_t$) would not be affected by the solar magnetic field lines, instead penetrating the solar surface directly. These particles would generate hadronic cascades in the Sun’s interior, and $\gamma$-rays would not be able to escape. Only CR particles with energy below $E_t$ would interact with the magnetic field configuration and be reflected by it. The cascades from these reflected CRs would eventually include $\gamma$-rays that escape the photosphere and heliosphere, and these are detectable as disk emission. \cite{Seckel-CR-Sun-1991} mention that the energy threshold $E_t$ varies depending on the solar magnetic field structure, strength, and scale under consideration, with values ranging from 300 GeV (for the `canopy, where flux tubes broaden out and merge') to 20 TeV (for the large-scale bipolar magnetic field).

The Fermi satellite has offered a unique glimpse into the $\gamma$-ray sky, enabling us to observe the behavior of the Sun's MeV-GeV emissions. While our understanding of solar $\gamma$-ray emission is still developing, current observations provide valuable data for models incorporating spatially and temporally resolved $\gamma$-ray fluxes. 

The recent detection of solar disk emissions at the TeV range by the High-Altitude Water Cherenkov \citep[HAWC;][]{HAWC-TeV-Sun-2023} expanded our observational knowledge and showed that solar TeV emissions anticorrelates with solar activity (i.e. a larger flux during solar minimum and a lower flux during solar maximum). In a previous work, \cite{KennyNg-GammaTvarSun-2016} used 6 yr of Fermi-LAT observations and were the first to report on the anticorrelation between solar activity and the GeV solar disk emission. The same modulation in the GeV band was later confirmed by \cite{TimLinden-2022} over an entire solar cycle (11 yr), in line with HAWC's measurements in the TeV band. \cite{HAWC-TeV-Sun-2023} observations with HAWC explores the energy range from approximately 500 GeV to 2.6 TeV, using a dataset collected from 2014 November to 2022 January. These observations suggest the corresponding spectral index to be 3.62$\pm$0.14, which is steeper than the CR spectrum. Given that the spectrum observed by Fermi-LAT in the GeV band is harder than the CR spectrum, HAWC data suggest a spectral break at approximately 400 GeV. This sets a crucial constraint for models aiming to reproduce solar emissions at very high energies.

\cite{Mazziotta-GammaSunSimFluka-2020} employed the FLUKA code \citep[a Monte Carlo code for simulating hadronic and electromagnetic interactions;][]{FLUKA-2014, FLUKA=2005} to simulate CR interactions with the Sun, incorporating a detailed account of the physical environment. The model used by \cite{Mazziotta-GammaSunSimFluka-2020} considers (i) Protons, helium, and electrons as primary CR components, with energy levels ranging from 100 MeV to 100 TeV; (ii) the development of CR cascades along the solar atmosphere, modeled using a density profile; (iii) the flux of CRs reaching the Sun, estimated based on the CR flux observed from Earth (as measured by AMS-02), and accounting for the propagation of CRs through the solar system up to the Sun (i.e., setting the flux at 1 au based on the CR flux at Earth); and (iv) a detailed magnetic field representation, using the Potential Field Source Surface (PFSS) model \citep{PFSS-Altshuler1969,PFSS-Schatten1969} and the \textit{Bifrost} stellar atmosphere simulation code \cite[][ with enhanced descriptions of solar flux tubes and the chromosphere]{Bifrost-Gudiksen2011}. According to \cite{Mazziotta-GammaSunSimFluka-2020}, protons, helium, and electrons contribute 74$\%$, 24$\%$, and 2$\%$ to the solar $\gamma$-ray emission, respectively.

Several other notable works have estimated solar disk $\gamma$-ray emissions, including \cite{ZheLi-GammaSunSim-Geant4-2020, ZheLi-GammaSimSunP2-2020} and \cite{Zhou}, which used the Geant4 toolkit \citep{Geant4-2003} for simulating particle propagation through matter. These simulations revealed that the primary channel for solar $\gamma$-ray production is the decay of neutral pions generated via proton-proton interactions (involving cosmic-ray protons and thermal protons in the solar plasma) or subsequent hadronic showers. A contribution of around 10$\%$ comes from electron and positron bremsstrahlung and positron annihilation, primarily during the final stages of secondary cascade showers. In addition, recent theoretical advancements by \cite{SmallScaleFields-Li2023} provide a model for small-scale magnetic fields and describe its contributions to shape the $\gamma$-ray spectral emission from the solar disk. In all cases mentioned, the observed flux for energies greater than 10 GeV deviates from the simulations and models, exhibiting a harder spectrum than expected.  

In a study with Fermi-LAT data, \cite{Tang-2018} reported a high-energy spectral signature in the solar $\gamma$-ray spectrum, resembling a spectral dip or flux gap in the 30-50 GeV band. The origin of this spectral feature remains unclear and deepens the mystery surrounding the solar disk $\gamma$-ray surplus (which can exceed the expected flux by  nearly an order of magnitude at the highest energies). Furthermore, recent simulations \citep{Mazziotta-GammaSunSimFluka-2020, ZheLi-GammaSunSim-Geant4-2020, ZheLi-GammaSimSunP2-2020} provide no insight into the reported spectral dip signatures at about 30-50 GeV.

A recent proposal by \cite{Solar-Tevatron-2023} posits that the Sun, during its quiet states, could function as a Tevatron, accelerating CRs up to TeV energies. In this model, protons confined within the chromosphere's magnetic field gain energy through shock acceleration (`acoustic-shock disturbances moving upward in the solar atmosphere'). Given the chromosphere's size, it can contain magnetically trapped TeV protons that interact with matter in the solar atmosphere, resulting in detectable $\gamma$-ray and neutrino signals. This approach can replicate the observed $\gamma$-ray spectrum continuum (including Fermi-LAT and HAWC data) and deserves further investigation. Yet, the puzzling spectral dip remains unexplained. 

In a previous work, \cite{TimLinden-2018} report on the solar disk $\gamma$-ray emission becoming ``more polar after the solar minimum" (of cycle 24) and that this trend increases significantly at higher energies. In that case, \cite{TimLinden-2018} studied solar disk samples with events before and after 2010 January (respectively covering 17 and 94 months each) and later in \cite{TimLinden-2022} incremented the study with data up to 2020. We now have the opportunity to extend their analysis, considering extra 3 yr of Fermi-LAT data and looking into different subsamples regarding solar activity, energy, and time. We look forward drawing a better understanding of the trends observed in past studies and to develop new data exploration \& visualization strategies.

Here in this work, we focus on the solar disk emission component, which dominates at GeV energies. We localize $\gamma$-ray events within helioprojective coordinates \citep{HelioCoordSyst}, tracking their occurrence in relation to the Sun's center and rotation axis. We further explore dependencies in connection with the energy band, solar cycles, and polar magnetic field strength. In selecting our $\gamma$-ray sample, we use a 5 GeV energy cut in order to restrict the analysis to events with high angular resolution. At this energy level, we also circumvent potential contamination from solar flares \citep{FermiLAT-SolarFlareCat-2021}. \cite[][their Figure 2]{TimLinden-2022} show the Fermi-LAT spectrum from 2008 to 2020, comparing between the entire dataset and a case where solar flares were removed, which mostly affects the $\gamma$-ray flux at E $<$ 1 GeV.

In addition, we take careful measures to account for the presence of foreground contaminants, including the Earth's limb, Sun-Moon encounters, and the presence of background high-energy/astroparticle sources, primarily: The third data release from the fourth Fermi Large Area Telescope catalog of $\gamma$-ray sources \citep[4FGL-DR3;][]{FermiLAT-4FGLDR3-2022, 4LAC-DR2}; The second catalog of flaring gamma-ray sources from the Fermi-LAT All-sky Variability analysis \citep[2FAV;][]{2FAV-2017}; Blazars and blazar-candidates from multiple catalogs \citep{5BZcat,masetti2013,1WHSP,2WHSP,3HSP,1BIGB,2BIGB}. In addition, we considered the first catalog of Fermi-LAT Long Term Transient sources \citep[1FLT;][]{1FLT-2021}, but found no match (in space and time) between the solar transits and the 1FLT transient events.

With this work, we study the $\gamma$-ray emission from the solar disk using model-independent statistical tests. We investigate the existence of anisotropic signatures in $\gamma$-ray emission and explore dependencies regarding energy and time.

\section{The Solar Disk Refined Sample}
\label{sec:sample-select}


This study investigates the distribution of $\gamma$-ray events associated with the solar disk. The data come from observations made by the Fermi-LAT satellite from 2008 August to 2022 January\footnote{For a comprehensive description of the Fermi-LAT mission, including a detailed discussion on instrument performance, calibration, and the associated ScienceTools, readers are invited to consult \cite{FermiLAT} and \cite{FermiTools:2019}}. Our event selection is carefully designed to account for any background events that could potentially taint the final sample. Such background events include Sun-Moon encounters, the Sun's transit across the galactic disk, proximity to 4FGL $\gamma$-ray perennial and transient sources, proximity to blazars and blazar candidates, and coincidences with $\gamma$-ray bursts (GRBs) listed in the Fermi-LAT $\gamma$-ray burst monitor (GBM) catalog.

The Fermi-LAT event files (FT1) provide essential information about each event such as energy, sky position (R.A. \& Dec. in J2000), time of occurrence, zenith angle, event-ID, event class, and event type. For the event class, the classification is based on the photon probability and the quality of the event reconstruction. Each event class is characterized by its own set of instrument response functions (IRFs). For the present analysis we have selected a sample of P8R3-SOURCE events (evclass=128), which are recommended for the analysis of moderately extended sources and point sources on medium to long timescales. The average point spread function (PSF) for these events is approximately 0.16$^\circ$ at 10 GeV, and asymptotically approaches 0.10$^\circ$ above 50 GeV.

Events within each class are further divided into different event types, according to selections based on event topology. In particular, in the Fermi-LAT PASS8 release \citep{PASS8}, each event is assigned a PSF event type, indicating the quality of the reconstructed direction. Events are divided into quartiles, designated as PSF0 (lowest-quality quartile, $evtype$=4), PSF1 ($evtype$=8), PSF2 ($evtype$=16) and PSF3 (highest quality quartile, $evtype$=32). While we follow up on PSF0 events, it's important to note that we do not use them in our analysis due to their relatively large PSF. The PSF 68\% containment radius varies with energy and event type. For instance, at 10 GeV, this value is approximately 0.17$^\circ$ for PSF1, 0.12$^\circ$ for PSF2, and 0.08$^\circ$ for PSF3, as described in the Fermi-LAT PASS8 documentation\footnote{Fermi-LAT PASS8 PSF: \url{https://www.slac.stanford.edu/exp/glast/groups/canda/lat_Performance.htm}}.

To follow up on the individual event's PSF, we use $gtselect$ to prepare a preselection and build four separate all-sky subsamples, respectively with PSF0 to PSF3\footnote{We use $gtselect$ with $evclass=128$ and $evtype$ corresponding to PSF0, PSF1, PSF2 and PSF3, to create four subsamples that separate events according to their PSF types.}. At this pre-selection stage, we recover all-sky events with energy exceeding 2.5 GeV, and with a maximum zenith angle of 110$^\circ$. For each subsample, we use $gtmktime$ to filter only good time intervals, with data quality assigned by the flags $(DATA\_QUAL>0)\&\&(LAT\_CONFIG==1)$. Throughout our sample selection, we will examine the use of different source-type events, focusing on events with the most accurately reconstructed positions. 

To accurately locate the Fermi-LAT observatory at the time of each event, and to track the positions of the Sun and Moon relative to it, we rely on additional information available in the spacecraft files (FT2) \footnote{Cicerone: \url{https://fermi.gsfc.nasa.gov/ssc/data/analysis/documentation/Cicerone/Cicerone_Data/LAT_DP.html}}. The Fermi-LAT FT2 files provide the Sun's position (RASUN, DECSUN [J2000]); however, the Moon's position needs separate computation and subsequent incorporation into the FT2 database. To accomplish this, we employed the Moonpos-1.2 tool\footnote{Moonpos-1.2, a user contribution tool, was authored by Dr. Paul Ray and has recently been updated: \url{https://fermi.gsfc.nasa.gov/ssc/data/analysis/user/}}, which factors in parallax correction to account for Fermi-LAT's continuous positional shift.

The FT2 files contain the spacecraft's position at 30 s intervals. We cross-matched the event files (FT1) with the spacecraft files (FT2), thereby assigning the spacecraft, Sun, and Moon positions at the time of each event. We employed a 30 s time bin and retained the closest match between FT1 event time and FT2 spacecraft time. With this data, we calculated the distance between the Sun and each $\gamma$-ray event within our sample. Additionally, we determined the distance between the Sun and the Moon at every time, so that we can account for Sun-Moon transits.

\begin{table*}
	\centering
	\caption{In this table we list the GRBs that happen close to the Sun. In the following columns, we specify the GRB name according to the Fermi GBM Burst Catalog (GRByymmddfff, i.e. year (y), month (m), day (d), fraction day (fff)), the  MJD time associated to the GRB event, the date, the R.A. and Dec. positions associated with the GRB, the solar position at the time (ra$_{\odot}$,dec$_{\odot}$), and the time windows removed (from T start to T end) in units of Mission Elapsed Time (MET).}
	\label{table:grb}
	{\def\arraystretch{1.1}
	\begin{tabular}{lllll|ll|ll} 
		\hline
	   GRB.ID  &  MJD  &  date  &  R.A.$_{grb}$ & Dec.$_{grb}$ & R.A.$_{\odot}$ & Dec.$_{\odot}$ & T start [MET]  &  T end [MET] \\
		\hline
    GRB100207721 & 55827.48 & 2010Feb07 & 321.78 & -15.78 & 321.112 & -15.225 &  287254160.6 &  287291960.6 \\ 
    GRB110923481 & 55827.48 & 2011Sep23 & 181.37 &  -1.60 & 179.941 &  0.0256 &  338468450.6 &  338506250.6 \\ 
    GRB121005030 & 56205.03 & 2012Oct05 & 195.17 &  -2.09 & 191.044 & -4.7472 &  371088849.6 &  371126649.6 \\ 
    GRB150705009 & 57208.00 & 2015Jul05 & 102.49 &  20.86 & 103.565 & 22.8517 &  457746018.6 &  457783818.6 \\ 
    GRB200922718 & 59114.72 & 2020Sep22 & 182.02 &  -1.96 & 179.883 &  0.0511 &  622485846.6 &  622523646.6 \\ 
        \hline
	\end{tabular}}
\end{table*}

\subsection{The base sample} 

We start our analysis with a selection of events within 0.5$^\circ$ of the solar center, with energies exceeding 5 GeV, zenith angles larger than 105$^\circ$, and PSF1-2-3 \footnote{Here, we select events from the pre-selection, as described in section sec.~\ref{sec:sample-select}}. This effectively eliminates PSF0 events associated with the poorest directional reconstruction and spatial uncertainty, enabling us to concentrate on solar disk emission. This preliminary selection yields 1125 events, which hereafter will be designated ``base sample."

The base sample selection emphasizes our focus on solar disk emission, which predominates over the IC component at GeV energy levels. Given the solar disk's angular radius of $ \rm r_\odot \approx 0.26^\circ$, the base sample encompasses an area approximately twice the size of the solar disk. The zenith-angle cut, following recommendations from \cite{3FHL-2017}, serves to exclude $\gamma$-ray events that might arise from CR interactions with the Earth's atmosphere.

\subsection{The clean sample: Foreground Cleanup}

In this section, we detail additional cuts to the base sample intended to minimize association of spurious events with the solar disk. We describe a sample selection that accounts for solar transits through the galactic disk, Sun-GRB coincidences, Sun-Moon encounters, and the solar transit through known (and candidate) $\gamma$-ray sources. As for the resulting selection, we refer to it as the ``clean sample." 

\textbf{Galactic Cut}: The galactic disk is a significant source of non-isotropic $\gamma$-ray foregrounds. To reduce this effect, we exclude 180 events from the base sample that occur while the sun transits through the galactic disk (at lower galactic latitude $|$b$|$ $<$ 10$^\circ$).

\textbf{Gamma-Ray Bursts (GRBs)}: Using the Fermi GBM Burst Catalog\footnote{The continuously updated GBM Burst Catalog, also referred to as `The FERMIGBRST database', is available at \url{https://heasarc.gsfc.nasa.gov/W3Browse/fermi/fermigbrst.html}.} \citep{GRBcat-2020}, we noted the dates and positions of listed GRBs, identifying any within 5.0$^\circ$ of the Sun. We then removed a time window of 30 minutes before and 10 hr after each GRB occurrence, as outlined in Table \ref{table:grb}. Despite the Solar-GRB encounters listed, this cut did not result in the removal of any events from the current sample (i.e. the base sample just after the galactic cut).

\textbf{Sun-Moon Transits:} The MoonposV1.2 tool provides reliable calculation of the Moon's positions (including parallax corrections) and was used to determine the Sun-Moon encounters. The Moon is known to emit $\gamma$-rays, a consequence of pion decay from CRs bombarding the lunar surface \citep{GammaMoon-Moskalenko2007ApJ,GammaMoon-2016}. The lunar $\gamma$-ray spectrum at the GeV band is relatively steep \citep{GammaMoon-2016, Johannesson-GammaSunMoon-2013} and faint in comparison to the solar disk GeV emission. Though it is unlikely to constitute a significant contaminant for the GeV component of solar disk emission, we removed all time windows related to Sun-Moon encounters within a distance of less than 2.0$^\circ$. This resulted in the removal of one event compared to the previous step.

\textbf{The Background of $\gamma$-ray Sources:} A source of contamination for the solar disk emission in the GeV-TeV band is the presence of $\gamma$-ray sources along the Sun’s path. We account for this by considering all sources in the 4FGL-DR3 catalog \citep{FermiLAT-4FGLDR3-2022} and the 2FAV list of flaring $\gamma$-ray sources \citep[which encompasses transients that could potentially contaminate our sample;][]{2FAV-2017}. Notably, sources known to exhibit extended $\gamma$-ray emission are flagged accordingly. We further include all blazars and robust blazar candidates from a series of catalogs \citep{5BZcat,1WHSP,2WHSP,1BIGB,3HSP,2BIGB,masetti2013}. Even if undetected in broad-time, broadband Fermi-LAT analysis, blazars and blazar candidates are potential $\gamma$-ray transients \citep[detectable only in short time windows,][]{Arsioli-LSP-2018}. This process results in a compilation of approximately 13.5 k entries, around 11 k of which are located at high galactic latitude $|$b$|$ $>$ 10$^\circ$ and may contaminate our sample in case of a solar transit.

On an event-by-event basis, we used the Astropy\footnote{Astropy: \url{http://www.astropy.org}} Python library \citep{astropy:2013, astropy:2018, astropy:2022} to compute the distance between the solar center and the nearest contaminant sources. We excluded all intervals where the Sun-to-source distance is $\leq$0.6$^{\circ}$. For extended sources, we implemented a larger Sun-to-source distance threshold of 1.0$^{\circ}$. This procedure resulted in the removal of 211 events.

Through this process, we ensure that even short-lived $\gamma$-ray transients (associated with blazars and blazar candidates) are accounted for on a case-by-case basis, and any potential contaminants are duly removed from our sample. We implemented all time and energy selection in a custom Python script, including cuts associated with the zenith angle of events, galactic latitude, Sun-Moon encounters, Sun-GRB encounters, Sun-Source encounters, and event type. The resulting clean sample contains 730 events with E$\geq$5 GeV.

\subsection{The Refined Sample: Quality of Event Reconstruction}
\label{sec:refined}

The main objective of our study is to inspect the localization of $\gamma$-ray photons that originate from the solar disk. In doing so, it is essential not only to build a clean sample but also to ensure that it includes only well-reconstructed events. Given that the solar radius (r$_{\odot}\approx$ 0.26$^\circ$) represents the characteristic angular scale of our system, we require that all events exhibit a characteristic PSF scale smaller than half the solar radius, i.e., 0.13$^\circ$. This PSF-based selection ensures a precise event localization that is crucial for our examination of the solar disk's $\gamma$-ray morphology, thereby bringing novel insights into its behavior.

\begin{table}
\centering
\caption{The first column presents the sub-sample short name, following the data selection described in sec. \ref{sec:sample-select}. The subsequent columns shows the number of events `n' with energies E$\geq$ 5, 8, 20, and 50 GeV. It should be noted that the energy levels of 5, 8, and 20 GeV are aligned with those used to construct the refined sample. An additional 50 GeV level is included to monitor the number of highest energy events.}
\label{table:sample-sizes}
{\def\arraystretch{1}
\begin{tabular}{l|llll} 
        & Number & of & events & with:      \\
subsample & E$\geq$5  & E$\geq$8  & E$\geq$20  & E$\geq$50 GeV \\   
        \hline
    Base sample     & 1125 & 628 & 180 & 52 \\
    Gal-cut         &  945 & 539 & 148 & 38 \\ 
    Sun-GRBs        &  945 & 539 & 148 & 38 \\ 
    Sun-Moon        &  944 & 538 & 147 & 38 \\
    Sun-sources     &  730 & 422 & 109 & 28 \\ 
    Refined sample  &  419 & 317 & 109 & 28 \\   
        \hline
 \end{tabular}}
\end{table}

Based on this consideration, we have implemented a further energy-dependent event selection to obtain a refined event sample. In the energy range 5-8 GeV we have retained only PSF3 events; in the energy range 8-20 GeV we have retained only PSF2 and PSF3 events; finally, for energies above 20 GeV we have retained only PSF1, PSF2 and PSF3 events. 

Table \ref{table:sample-sizes} presents the event count for the base sample to the refined sample. We examine the sample sizes at each selection stage, counting all events with energy exceeding 5, 8, 20, and 50 GeV. Note, the base sample includes events occurring within 0.5$^\circ$ of the sun center, with E $>$ 5 GeV, zenith angle$\leq$105$^\circ$, and PSF1-2-3. Subsequent selection steps aim to cleanse the sample, with the final stage focusing on refining it to include only the best reconstructed events.

When considering all events with PSF1-2-3, the primary reason for the drop in the number of events in the energy range between 5 and 20 GeV is the stringent quality of reconstruction requirements set by the refined sample, leading to a reduction of 43\%-25\%, respectively. Next, the galactic cut significantly impacts all energy ranges, removing between 15\% and 25\% of the base sample events, from the lowest to highest energies. The process of accounting for solar encounters with detected and potential $\gamma$-ray sources is another crucial step in refining the sample, which results in the elimination of approximately 25\% of events across all energy ranges. The cuts related to GRBs and Moon transits have the least impact on the final sample size.

The data presented in Table \ref{table:sample-sizes} are noteworthy for four key reasons: (i) the potential contamination from solar encounters with $\gamma$-ray sources and transients is significant, and accounting for those encounters lead to the removal of approximately 25\% of the events; (ii) at energy levels $\geq$5 GeV and outside the galactic disk, the contamination of solar disk emission due to GRBs and Moon transits is relatively low; (iii) the solar disk remains a significant emitter at E $>$ 20 GeV, even after all data cuts, and this holds true for the best event types 16 and 32.

\subsection{The VHE Events at E$>$90 GeV} 

In Table \ref{tab:VHE} we list all events with E$\geq$90 GeV from a cone with 0.5$^\circ$ angular aperture centered on the Sun   and with a zenith angle of up to 110$^\circ$. We take into account all events types (i.e. PSF0 to PSF3) and monitor the point in our selection at which each event is eliminated.

We identified a total of 16 events with E $>$ 90 GeV, and more than a third survived the selection steps that led to the clean sample. Removed events (and its corresponding numbers) include: three PSF0 events, another four occurred during solar transits through the galactic disk (at $|$b$|$$\leq$ 10$^{\circ}$), two were removed due to close proximity with potential $\gamma$-ray emitters, and one had a zenith angle of $\approx$106$^{\circ}$. The events removed due to close proximity with sources involved a confirmed blazar 5BZQ J0848+1835, and the blazar candidate CRATES J041840+211632. Despite implementing stringent sample refinement criteria, six events with E $>$ 90 GeV survived, with the highest-energy event of 138.4 GeV. Our analysis confirms the solar disk as a very-high-energy (VHE) emitter, extending the findings from \cite{TimLinden-2022} and in line with recent measurements from HAWC \citep{HAWC-TeV-Sun-2023}.

\section{Localization of Solar Disk $\gamma$-Rays: Helioprojection}
\label{sec:loc}

Our refined sample now holds the best reconstructed events (all with characteristic PSF smaller than $\approx$0.13$^\circ$ and E $>$ 5 GeV) out of the galactic disk $|$b$|$ $>$ 10$^\circ$, and that survived the sample selection regarding solar close encounter with the Moon, GBRs and $\gamma$-ray sources/transients/candidates. We use the Python library SunPy\footnote{SunPy: \url{https://sunpy.org/}} \citep{SunPy-Mumford2020} to convert the events positions into a helioprojection \citep{HelioCoordSyst}. SunPy includes parallax correction due to the observer's motion with respect to the observed target (which causes apparent displacement of the target when viewed from different positions).

The helioprojection converts the positions of events observed with Fermi-LAT (R.A., Dec. in J2000) into a reference frame with origin at the solar disk centre, and with axis directed along the Sun's rotation axes (Ty) and perpendicular to it (Tx). In sec. \ref{sec:stat-test}, we describe model-independent statistical tests considering subsamples with the entire dataset (from 2008 August to 2022 January), but also for subsamples corresponding to time intervals centered on the maximum (max) and minimum (min) solar activity. Based on considerations regarding the CR spectrum and the solar activity \cite[sunspot number;][]{sun-polar-cycle25-2023}, we define the following periods associated with solar maxima and minima:

\begin{table*}
\caption{This table provides a list of all events with E $>$ 90 GeV that occur within 0.5$^\circ$ of the Sun and have a zenith angle up to 110$^{\circ}$. The upper section of the table enumerates events that survived our selection process, while the lower section lists the events that were removed due to their type (PSF), their location within the galactic plane (gal-cut), their proximity to a blazar (bz) or a blazar candidate (bz-cand), or their zenith angle being greater than 105$^\circ$.}
\centering
\begin{tabular}{l|rrrrrrrrr}
\toprule
E[GeV] & PSF & R.A. & Dec. &  B & Zenith &  Dist-Sun[deg] & Event.ID & Time[MET] & removed \\
\hline
117.2 & PSF1 &   1.336721 &   0.703271 & -60.0 & 40.4 & 0.254 &  6168824 & 2.594042e+08 & -  \\
112.6 & PSF1 & 235.905228 & -19.473272 &  27.4 & 54.2 & 0.287 & 13163357 & 2.803965e+08 & -  \\
 95.7 & PSF2 & 302.758575 & -20.092472 & -26.1 & 44.1 & 0.148 &  1665684 & 2.541428e+08 & -  \\
138.4 & PSF2 & 144.415894 &  14.299705 &  43.2 & 59.2 & 0.260 &  7244859 & 2.719917e+08 & -  \\
 92.9 & PSF2 & 228.499466 & -18.012501 &  33.1 & 28.2 & 0.081 &  6772459 & 2.797988e+08 & -  \\
 94.5 & PSF3 & 161.036316 &   8.175216 &  54.4 & 57.6 & 0.148 &  7193764 & 3.366444e+08 & -  \\
\hline
\hline 
212.8 &  PSF0 & 224.497284 & -16.850988 &  36.3 & 47.4 & 0.067 &  7567307 & 2.478953e+08 & PSF      \\
103.2 &  PSF0 & 260.345856 & -23.102371 &   7.7 & 76.9 & 0.398 &  5095854 & 2.508446e+08 & PSF      \\
162.3 &  PSF0 & 326.701416 & -13.468830 & -44.8 & 70.5 & 0.360 &  7285392 & 5.402370e+08 & PSF      \\
467.6 &  PSF1 & 268.045502 & -23.177397 &   1.7 & 52.9 & 0.337 &  3623019 & 2.829892e+08 & gal-cut  \\
226.8 &  PSF1 & 272.899139 & -23.342749 &  -2.2 & 21.7 & 0.068 &  5799030 & 2.517901e+08 & gal-cut  \\
126.1 &  PSF1 & 265.657532 & -23.281782 &   3.5 & 99.0 & 0.105 &   372387 & 5.036870e+08 & gal-cut  \\
139.3 &  PSF2 & 260.706970 & -23.243065 &   7.3 & 55.5 & 0.126 &  3030273 & 2.508316e+08 & gal-cut  \\
320.0 &  PSF1 &  64.301765 &  21.224283 & -20.6 & 36.6 & 0.254 &  7858649 & 6.438328e+08 & bz-cand  \\
 97.6 & PSF2 & 130.834869 &  18.362808 &  32.6 &  5.9 & 0.195 &  1057159 & 3.022988e+08 & bz        \\
150.6 & PSF2 & 223.126266 & -16.625463 &  37.3 &106.2 & 0.269 & 12063477 & 5.002214e+08 & zenith    \\
\end{tabular}
\label{tab:VHE}
\end{table*}

\begin{itemize}
    \item First minimum (Min1): from 2008 August to 2011 July (MET 239557417.0-331171202.0)
    \item First maximum (Max1): from 2011 July to 2016 January (MET 331171202.0-473299204.0)
    \item Second minimum (Min2): from 2016 January to 2022 January (MET 473299204.0-662774405.0)
\end{itemize}

For the data analysis, we merge the two periods associated with solar minimum (Min1 + Min2), which we refer to as ``min1p2" along the text and plot legends.

\subsection{A look into the Radial Distribution of Events} 
\label{sec:radial-dist}

In this section, we present the radial distribution of events for the Min1p2 and Max1 samples. We study the events distribution considering all events in the refined sample, which have energies ranging from 5 GeV to 140 GeV and a characteristic PSF of approximately $\approx$0.13$^\circ$ or better.

\begin{figure}
    \centering
    \includegraphics[width=1.0\linewidth]{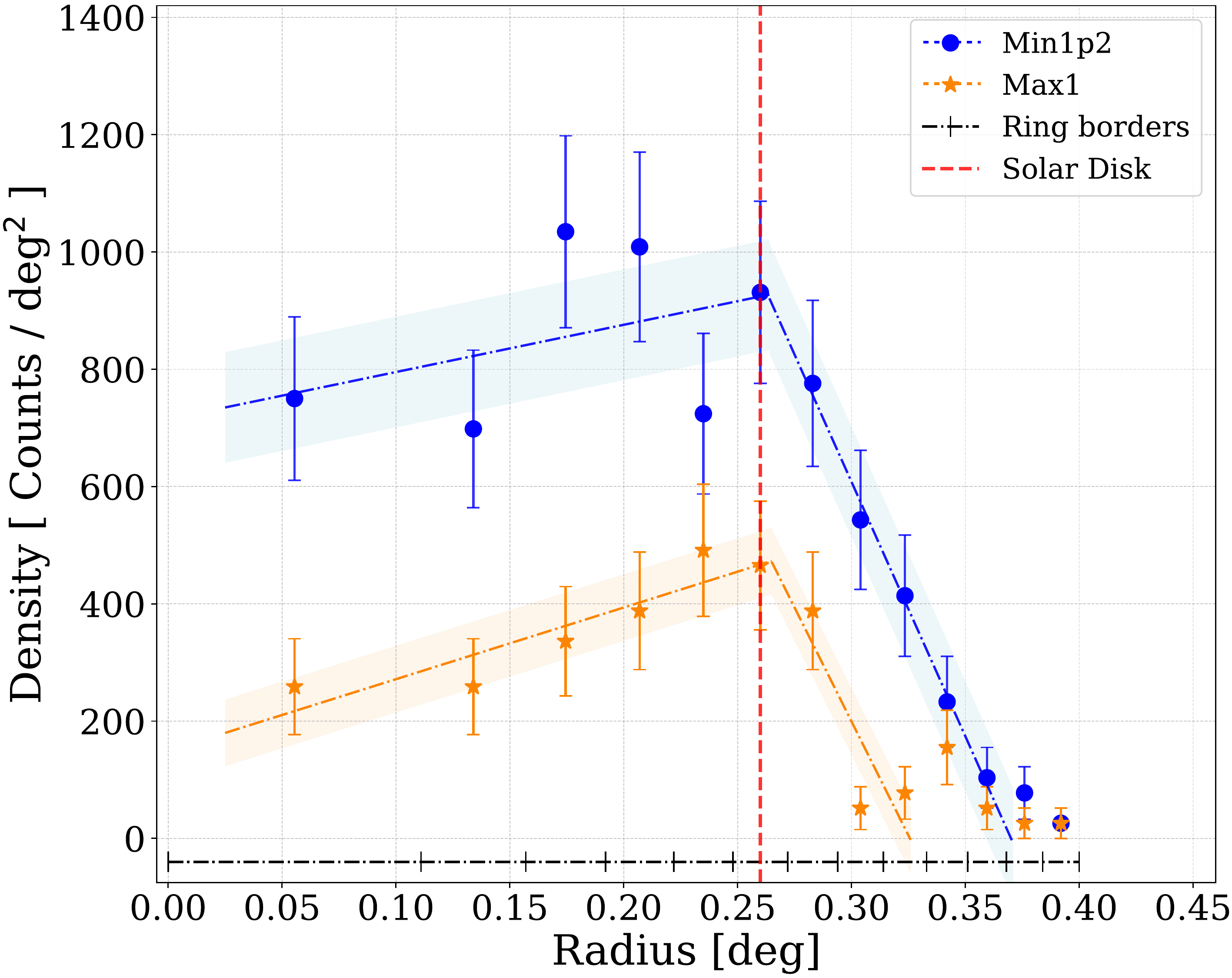}
    \caption{The radial distribution of events for the Min1p2 (blue) and Max1 (orange) samples, considering all events with E $>$ 5 GeV. We highlight the solar disk angular size with a red dashed vertical line. In the bottom of the plot, we show the border of the rings (bins) of constant area, over which we count events. The bins are marked with `$|$' over the black dashed horizontal line. The blue (orange) dashed line represents a broken linear fit (blf) to the radial density of events for the Min1p2 (Max1) sample, with its corresponding 1$\sigma$ uncertainty.}
    \label{fig:rad-evt}
\end{figure}

We select a region around the solar disk, extending from the center to r$_{out}$=0.4$^\circ$, with a total area of $\pi r^2_{out}$. We divide this region into $\rm n_i$=13 rings of equal area $\pi r^2_{out} / n_i$. We measure the density of events along the solar radius by counting the number of events that occur within each ring and divide by the ring area. The error bars for the density are calculated as the square root of the counts divided by the ring area; here we assume that the counts in each ring are Poisson distributed. In Figure \ref{fig:rad-evt}, the bins (i.e., the ring borders) are highlighted with the marker ``$|$" over the dotted-dashed black line. 

The radial distribution of events for the Min1p2 and Max1 samples, as shown in Figure \ref{fig:rad-evt}, provides insightful observations. Fitting the E $>$ 5 GeV events with a broken linear function, we measure the break (r$_{break}$) at approximately 0.264$^\circ$ and 0.265$^\circ$ for the Min1p2 and Max1 samples, respectively. These measurements are consistent with the angular size of the solar disk ($\approx$0.26$^\circ$). The fact that the break corresponds closely to the solar disk's angular size is a significant observation. It suggests that the emission in our refined sample is predominantly from the solar disk. If a substantial contamination were present, we would not expect the r$_{break}$ to align with the solar disk's angular size.

In both the Max1 and Min1p2 samples, the density of events exhibits a sharp decay just outside the solar disk, reaching a relatively low density within 0.1$^{\circ}$ from the disk. If the E $>$ 5 GeV emission is primarily from the solar disk, a steep decrease in event counts is anticipated beyond the solar disk, which aligns with our observations. Given that the refined sample only includes events with a characteristic PSF of $\leq$0.13$^{\circ}$, this sharp decay occurs within an angular scale that is consistent with the PSF size.

\subsection{A Look Into the Number of Events per E$_{bin}$}
\label{sec:Ebin}
  
To investigate the $\gamma$-ray emission from the solar disk, we use model-independent statistical tests to compare the distribution of events in Tx and Ty, and to examine the uniformity of their circular distribution. We have performed these tests on the whole dataset as well as the subsamples corresponding to the maxima and minima of the solar activity. We investigate the energy range between 5 to 150 GeV, allowing us to discern energy-dependent trends.

\begin{figure}
    \centering
    \includegraphics[width=1.0\linewidth]{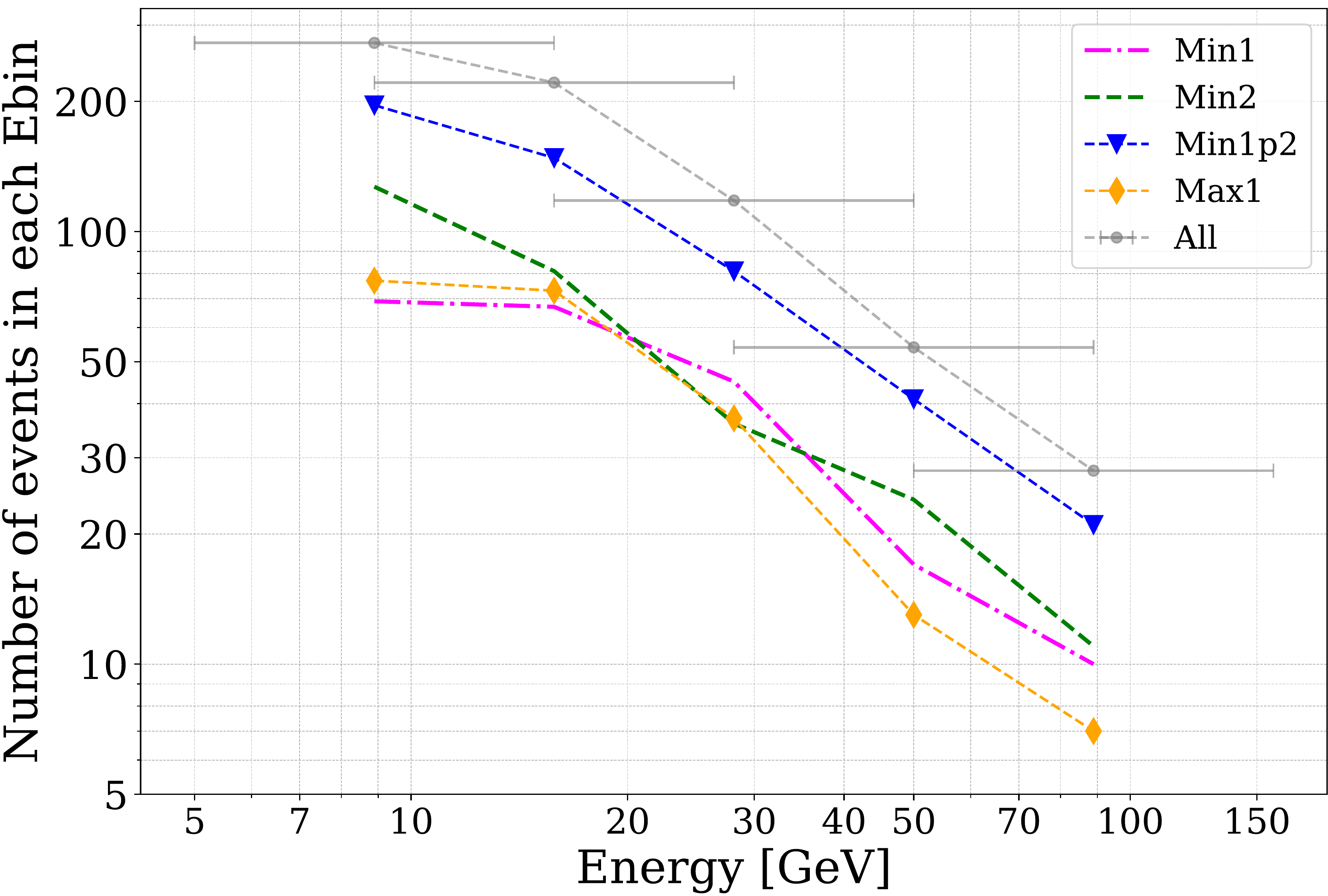}
    \caption{This figure represents the count of events for each energy bin. Note: this does not correspond to an intrinsic flux spectra from the Sun because the energy-dependent effects of exposure have not been taken into account. The energy bins maintain a constant size in logarithmic scale, equivalent to 0.5 dex. As a point of reference, the bin size is represented by the x-axis error bars for the data points labeled as ``All." The other labels maintain the same energy-bin size for each of their corresponding data points. The magenta dotted-dashed and the green dashed lines refer to the Min1 and Min2 subsamples, respectively. The blue triangle marker represents the Min1p2 subsample, while the orange diamond marker represents the Max1 subsample. Lastly, the grey dashed line (with circle markers) refers to the entire sample encompassing ``All" events, Max1+Min1p2.}
    \label{fig:n-events}
\end{figure}

We define our energy bins in a logarithmic scale, each spanning a width of 0.5 dex \citep{Allen-dex-1951}. To ensure comprehensive coverage and reduce potential biases, we overlap these bins by 0.25 dex. This means our bins cover energy ranges of approximately 10$^{0.7}$-10$^{1.2}$ GeV, 10$^{0.95}$-10$^{1.45}$ GeV, and so on, with a total of five energy bins. This overlapping approach enhances data visualization and minimizes biases tied to the choice of the starting energy for the bins. Figure \ref{fig:n-events} shows the number of events for each subsample and energy bin of the refined sample. It is meant to help visualize the number of events used in statistical tests that rely on the energy bins and does not incorporate exposure corrections\footnote{Therefore, Figure \ref{fig:n-events} does not represent spectral information of any kind (such as the $\gamma$-ray flux).}.

\section{Model-Independent Tests: Investigating Non-isotropic emission}
\label{sec:stat-test}

In this section, we describe our model-independent strategy for assessing signatures of anisotropic $\gamma$-ray emission from the solar disk. For instance, presuming isotropic emission requires a detailed model of the solar disk spectra. The spectrum dependence on energy determines the counts at a given energy bin; in turn, these counts are distributed according to the PSF and the underlying solar disk geometry. Summing these distributions over a given energy range would lead to a biased expected distribution of events because the spectrum bears inherent uncertainties. Therefore, to investigate the solar disk emission, we focus on model-independent tests.

We analyze our dataset using a combination of the Kolmogorov-Smirnov (KS) test -a non-parametric test of the equality of continuous, one-dimensional probability distributions\citep{KStest-FMassey-1978,KStest-Darling-1957}- and the Rayleigh Test -a statistical test for circular distributed data \citep{Fisher-SphereStat-1953, Berens-CircStat-2009}. These tests offer a robust and versatile tool to examine the geometry of the $\gamma$-ray emissions and detect deviations from isotropy, without the constrains from pre-defined models.

\subsection{Kolmogorov-Smirnov KS test} 
\label{sec:ks}

In order to investigate potential anisotropy in the $\gamma$-ray emission from the solar disk, we use the KS test. This test allows for a model-independent comparison of the Tx and Ty distributions of the $\gamma$-ray events. As there is no reason for Tx and Ty to mirror each other's distributions (i.e., the distributions are independent), we can leverage this property to detect signs of non-isotropic emission. 

The KS test compares the cumulative distribution function (CDF) of two samples, yielding a p-value that represents the likelihood that the two samples originate from the same underlying distribution. Conventionally, a p-value threshold of 0.05 is used to reject the null hypothesis, meaning the Tx and Ty distributions differ. In our exploratory analysis, we evaluate the statistical tests over five energy bins (\textit{n} = 5) within each of the independent subsamples, such as Max1 and Min1p2. To mitigate the risk of type I errors (false positives), we implement the Bonferroni correction, a well-regarded and conservative method to adjust p-values in scenarios with multiple tests \citep{bonferroni1936, Hochberg-pvalue-1988}. We apply the Bonferroni correction separately to each independent subsample and adjust the p-value threshold to 0.05/\textit{n} = 0.01. We consider p-values falling below this adjusted threshold as indicative of significant anisotropic trends in the $\gamma$-ray emission from the solar disk.

\begin{figure}
    \centering
    \includegraphics[width=1.0\linewidth]{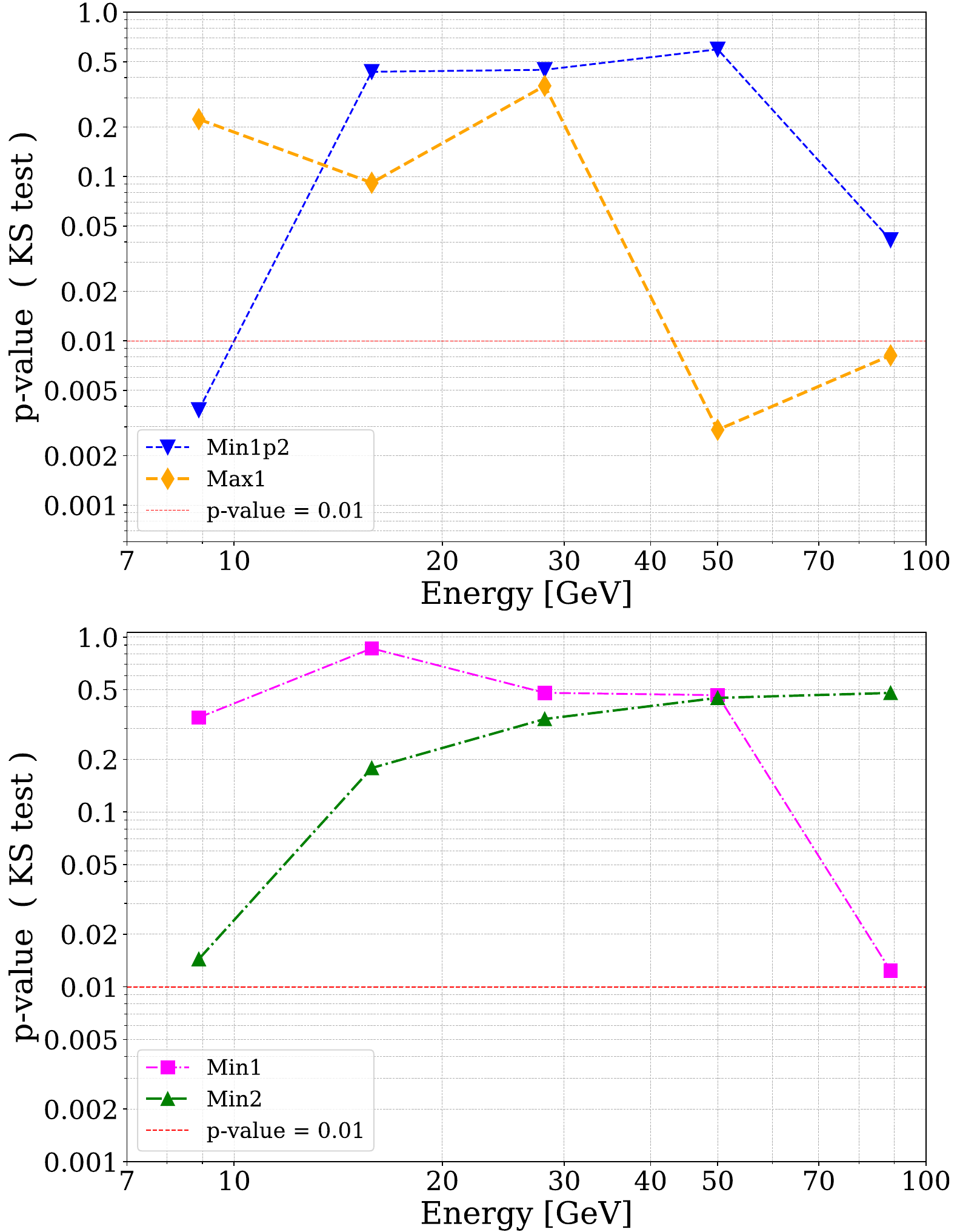}
    \caption{KS test p-values for each energy bin, which have a constant size in log-scale (0.5 dex), as shown in Figure \ref{fig:n-events}. The red dashed line marks the p-value threshold of 0.01, below which anisotropic emission is indicated. The upper plot displays the Tx vs. Ty comparisons for the Min1p2 (blue triangle markers) and Max1 (orange diamond markers) subsamples. In the lower plot, Tx vs. Ty comparisons for the Min1 (dotted-dashed magenta line with square markers) and Min2 (dotted-dashed green line with triangle markers) subsamples are shown.}
    \label{fig:ks-test}
\end{figure}

For the KS test, we use the $KS\_2samp$ function from SciPy.stats Python library \citep{SciPy-2020}, specifying the two-sided test. This approach evaluates the two samples irrespective of the direction of the difference and is sensitive to any discrepancy between distributions. In the event of anisotropy in the Tx and Ty distributions, a significant disparity between the CDF of the two samples would emerge, as determined by the KS test. It's crucial to note, however, that while the KS test can identify anisotropic trends, it does not specify their direction or nature. We feed the $KS\_2samp$ test with Tx and Ty values for each energy bin and for the corresponding subsamples: Min1 + Min2 (which we call Min1p2), Max1, and separately investigate the data associated to Min1 and Min2. Figure \ref{fig:ks-test} shows the p-values resulting from the KS test for the Tx and Ty comparisons.

\subsection{The Rayleigh Test} 
\label{sec:ray}

\begin{figure}
    \centering
    \includegraphics[width=1.0\linewidth]{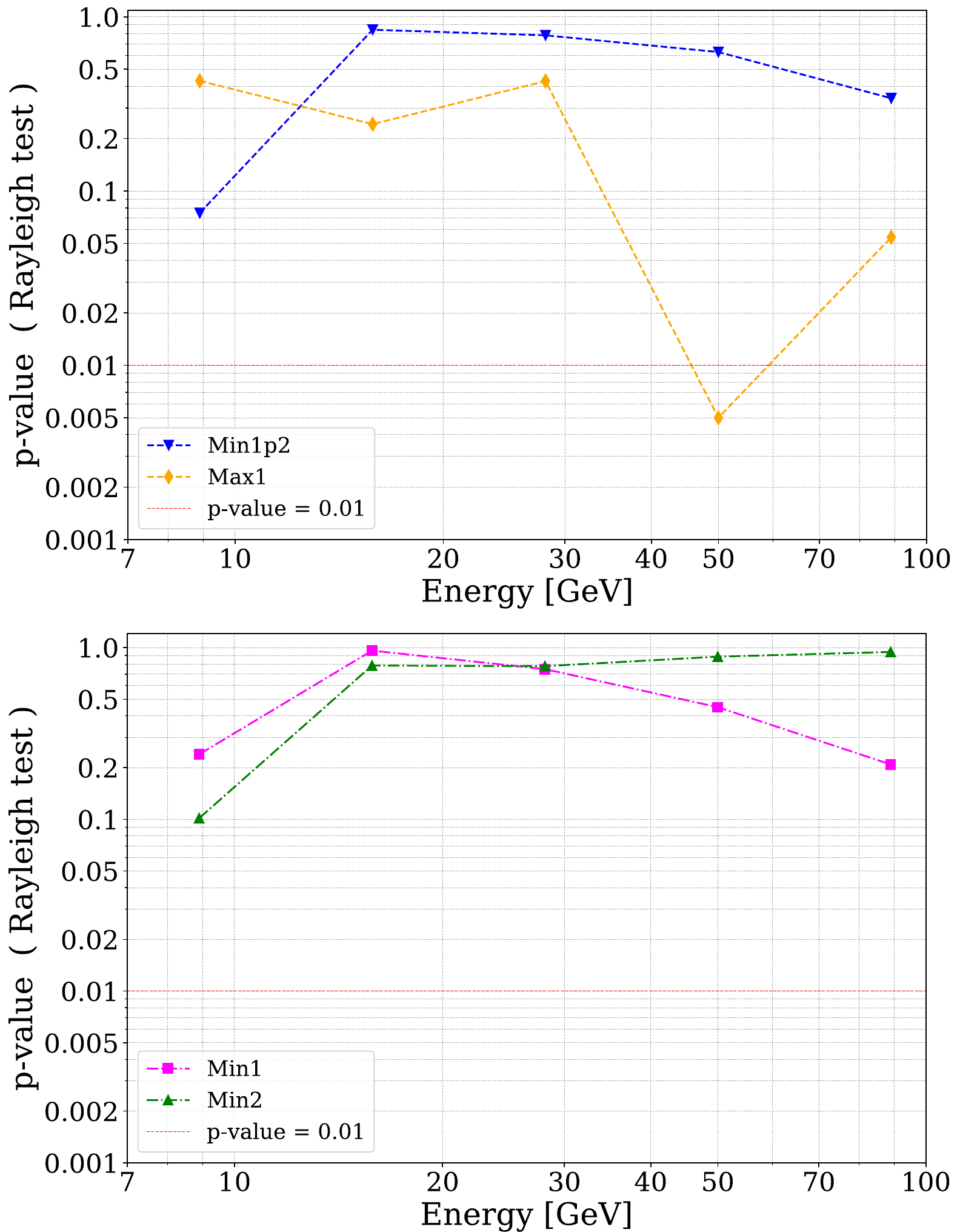}
    \caption{Rayleigh p-values corresponding to each energy bin. The energy bins have a constant size in log-scale (corresponding to 0.5 dex), just as in Figure \ref{fig:n-events}. A red dashed line highlight the p-value threshold of 0.01, below which nonuniformity is indicated. In the upper plot, the blue triangle marker refer to the Min1p2 subsample, while the orange diamond marker indicates the Max1 subsample. In the lower plot, dotted-dashed lines in magenta and green (with square and triangle markers, respectively) refers to the Min1 and Min2 subsamples.}
    \label{fig:rayleigh-test}
\end{figure}

The Rayleigh test is a powerful tool derived from circular statistics \citep{Fisher-SphereStat-1953}, employed to determine if a circular distribution is symmetric or exhibits deviations from uniformity. The Rayleigh test considers only the observed distribution of event angles. The test procedure involves (i) projecting the events onto a unity-radius sphere to define event-vectors; (ii) summing the vectors and normalizing the result vector by the number of event-vectors used. In the ideal case of a uniform distribution, the resulting vector would be close to null. Therefore, the p-value indicates how likely the resulting vector deviates from the ideal case. This test is particularly powerful in detecting unidirectional modality \citep{Berens-CircStat-2009, Fisher-CircStat-1996} and essentially measures the concentration of angles around a mean direction. In contrast, dipole or quadrupole structures in the data distribution would be harder for the Rayleigh test to resolve. As \cite{RayleighTest-Brazier-1994} demonstrates, the Rayleigh test is sensitive to small deviations from uniformity in circular data, even for relatively small samples. To implement the Rayleigh test, we first calculate the events' positions in polar coordinates, assigning a radius and angle to each solar disk event. We then use the $astropy.rayleightest$ library\footnote{Astropy.Rayleigh: \url{https://docs.astropy.org/en/stable/api/astropy.stats.rayleightest.html}} to evaluate the distribution of event angles, and Figure \ref{fig:rayleigh-test} shows the p-values resulting from the Rayleigh test.

\subsection{Visualization of the Events' Localization, Density and Distributions}
\label{sec:diagnostics}

\begin{figure*}
    \centering
    \includegraphics[width=0.8\linewidth]{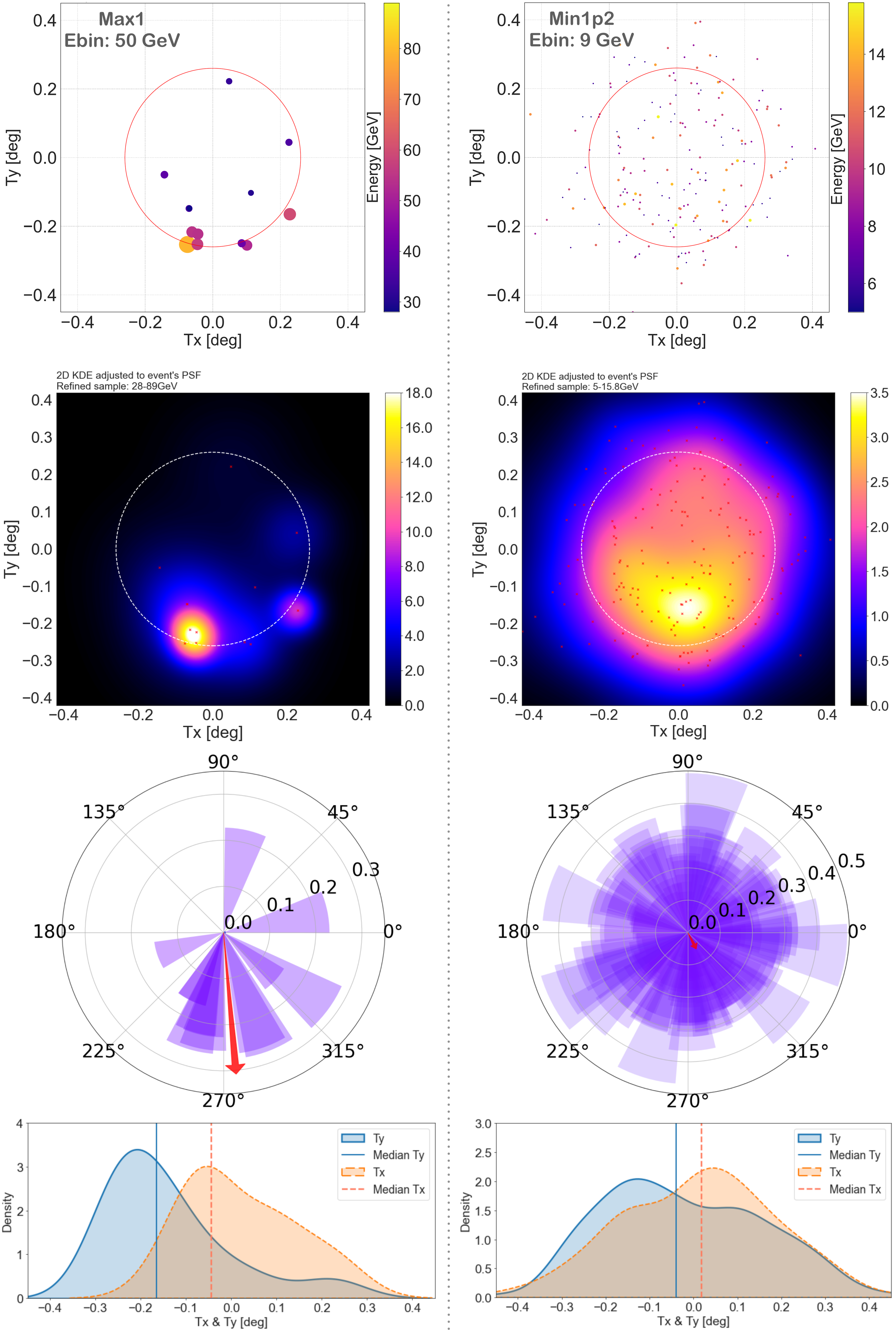}
    \caption{This figure presents two sets of images with visualization diagnostics for the anisotropic trends seen for the Max1 sample at 50-90 GeV energy bins (left); and for the Min1p2 sample at the 9 GeV energy bin. From top to bottom, the figure shows (i) the localization of events over the solar disk, (ii) the 2D KDE density distribution, (iii) the circular distribution of events, and (vi) the Tx \& Ty histograms. All visualizations are described in sec. \ref{sec:diagnostics}.}
    \label{fig:diagnostics}
\end{figure*}

The KS and Rayleigh tests showed the presence of significant anisotropic trends, with p-vales going below 0.01 for the energy bins at 9 GeV (Min1p2) and 50-90 GeV (Max1). To better investigate and visualize those statistical results, we implement a series of diagnostic plots (Figure \ref{fig:diagnostics}), which we describe in this section.

\textbf{The Localization of events.} In the first row of Figure \ref{fig:diagnostics} we depict the localization of individual events across the solar disk. In each plot we employ color-coding to represent the energy levels (as defined in the color bar). For easy differentiation between the lowest to highest-energy events, the marker size reflects the magnitude of the event's energy along all plots (i.e. with `small to large markers' covering the energy range between 5 and 130 GeV).

\textbf{The 2D KDE.} To better explore and visualize the results from KS and Rayleigh tests, we implement a kernel density estimate (KDE) of the event distribution. Instead of using standard KDE functions, we employ a custom 2D KDE approach that allows for varying bandwidths based on each event's PSF. Our KDE approach computes the density representation by assigning a Gaussian function to each event, centered at its position. The width $\sigma$ of the Gaussian function is determined by the PSF associated with each event, allowing for a representation of the event's density that accounts for the event's reconstruction uncertainties (e.g. the 2D KDE, Figure \ref{fig:diagnostics}). We use the $multivariate\_normal.pdf$ probability density function from SciPy.stats Python library \citep{SciPy-2020} to perform the 2D KDE according to the equation: 

{\small
\begin{equation}
\text{KDE}_{(x, y)} = \frac{1}{N} \sum_{i=1}^{N} \frac{1}{2\pi\sigma_{x,i}\sigma_{y,i}} \exp\left(-\frac{(x - x_i)^2}{2\sigma_{x,i}^2} - \frac{(y - y_i)^2}{2\sigma_{y,i}^2}\right)
\end{equation}} 
where $N$ is the number of events, ($x_i$, $y_i$) are the (Tx, Ty) coordinates of each event, $\sigma_{x,i}$ and $\sigma_{y,i}$ are the PSF scales of the i-th event along x and y axis. We assume a symmetric PSF, therefore $\sigma_{x,i}$=$\sigma_{y,i}$.   

In this representation, events with smaller PSF values -indicating better localization- contribute more heavily to the density at their respective positions. In contrast, events with larger PSF values have their contributions more diffusely spread, effectively diluting their impact on the overall density. For reference, the position of individual events is marked as small red crosses. The sum of all Gaussian functions is normalized such that its integral over the entire volume equals unity. This facilitates the comparison of event concentrations across different time windows and regions, as higher values on the color bar indicate more concentrated distributions of events.

\textbf{The Circular Distribution of Events.} In Figure \ref{fig:diagnostics}, third row, we display the circular distribution of the events over the solar disk, where each purple bar represent a single event and the bar length represents its radial position. This plot also shows the `trigonometric moment arrow' (depicted in red), which points toward a preferential direction of the  event's distribution. This arrow is calculated as the sum of unit vectors, under the assumption that the events are distributed over a circle with unitary radius. For the purposes of Figure \ref{fig:diagnostics} (and the following), we normalize the resulting vector assuming a sphere with a radius of R = 0.5$^{\circ}$, which better suits the angular scale of our case and offers a more reliable visualization. The arrow's direction represents the mean angle of the circular distribution, and its length indicates the degree of departure from a circular uniform distribution. For the circular statistics, we employed the $astropy.stats.circstats$ package\footnote{Astropy Circular Statistics: \url{https://docs.astropy.org/en/stable/stats/circ.html}}.

\textbf{Tx \& Ty Histograms.} At the bottom of Figure \ref{fig:diagnostics} we show the distribution of events in Tx and Ty coordinate space, respectively in shaded orange and blue. This representation illustrates the concentration of events regarding the Sun's rotation axis (Ty), and perpendicular to it (Tx), together with the corresponding median values for each distribution. Intuitively, anisotropies driven by Ty distribution shifts could be connected to long-lasting patterns of the solar polar field, while Tx-driven anisotropies would be more challenging to explain given the Sun's rotation (i.e. the constant shift of true physical positions of the solar surface). We should note that the complexities involved in modeling CR flux impacting the solar surface are intensively explored but not fully resolved in current models \citep{Mazziotta-GammaSunSimFluka-2020, ZheLi-GammaSunSim-Geant4-2020, ZheLi-GammaSimSunP2-2020, SmallScaleFields-Li2023}, e.g. variations of the CR flux bombarding the solar atmosphere could play a relevant role. These models, while advanced, do not fully account for the observed fluxes above 10 GeV, nor explain the 30-50GeV spectral dip \citep{Tang-2018}. Therefore, investigating Tx anisotropic hints is indeed relevant and might offer new insights into the intricate interactions of CRs and the solar atmosphere. Currently, it is challenging because with Fermi-LAT we need large time windows to collect solar $\gamma$-ray events at E $>$ 5 GeV, potentially washing out Tx-driven anisotropy signatures.

\section{Discussion: Results from KS and Rayleigh Tests}

Here we discuss the results from the KS and Rayleigh tests, which were applied to subsamples in energy and which cover time windows corresponding to the solar cycles Max1 and Min1p2, as defined in sec. \ref{sec:loc}. 

\subsection{The anisotropic trend at the Max1 highest energies} 

The Rayleigh test reveals a substantial deviation from circular uniformity for the Max1 subsample at the energy channel of 50 GeV, which translates to a preferential direction -unidirectional excess- in the circular distribution. Here, the p-value falls down to 0.005 level (Figure \ref{fig:rayleigh-test}) suggesting a significant anisotropic trend in the $\gamma$-ray emission. For this same channel, the KS test presents p-value of 0.003 (Figure \ref{fig:ks-test}) and indicates the Tx and Ty distributions are different,  reinforcing the evidence of $\gamma$-ray anisotropy.   

For the Max1 sample, the trigonometric moment arrow (depicted in red, Figure \ref{fig:diagnostics}) points toward a preferential direction, the solar south pole. The presence of anisotropic trends in both KS and Rayleigh tests provides a compelling signature of anisotropic $\gamma$-ray emission from the solar disk. Particularly, the 50 GeV Max1 channel has a robust sample of 13 well localized events with energy between 28 and 89 GeV. In Table \ref{tab:E50}, we present the Max1 events which concentrate in the solar south pole (the 2D KDE plot, Figure \ref{fig:diagnostics}).

\begin{table}
\caption{A list of E $>$ 50 GeV events associated with the high-density spot in the Max1 sample, located at Ty $<$ -0.1.}      \centering
\begin{tabular}{rrrrrr} 
\toprule
Time[MJD]  &  Energy[GeV] &  PSF &  Tx & Ty \\
\hline
55784.128  & 52.2 & PSF1 &  0.1010 & -0.2555 \\
55815.187  & 53.8 & PSF1 & -0.0441 & -0.2227 \\
56055.945  & 56.7 & PSF3 & -0.0454 & -0.2525 \\
56562.054  & 54.7 & PSF3 & -0.0615 & -0.2167 \\
56579.748  & 79.1 & PSF1 & -0.0749 & -0.2531 \\
57138.580  & 59.7 & PSF3 &  0.2279 & -0.1650 \\
\end{tabular}
\label{tab:E50}
\end{table}

The events associated with this high-density spot, despite being spatially concentrated, were recorded at intervals of at least 15 days, spanning nearly 750 days (between MJD 55815 to 56579; i.e. from 2011 September 1; to 2013 October 14). In terms of solar activity, this period coincides with an unusual polarity flip in the Sun \citep{sun-polar-Janardhan-2018, sun-polar-cycle25-2023}, with the south polarity flipping in 2013 November, and the north polarity flip delayed by a prolonged period of nearly null radial polar magnetic field strength from 2012 June to 2014 November.

In previous observations of solar phenomena, \cite{sun-polar-Petrie-2014} pointed out the asymmetric solar activity of the north and south poles during the transition between cycle 23 and 24, with the southern hemisphere being the most active during 2014 (during the decline of cycle 23). Therefore, asymmetry and anisotropy are common in solar physics. \cite{sun-polar-Petrie-2014} also note that these anisotropic features are often overlooked in flux transport models, which may weaken and compromise predictions of the solar disk $\gamma$-ray emission. Our findings echo the known anisotropic solar activity and raise new questions about the mechanisms driving the solar $\gamma$-ray behavior.

\subsection{The Min1p2 Anisotropy at the 9 GeV Energy Bin}  

For the Tx vs. Ty comparison of the Min1p2 sample, the KS test showed p-values of 0.004 at the 9 GeV E$_{bin}$, indicating an anisotropic trend of the solar disk $\gamma$-ray emission. Nonetheless, the Rayleigh test does not reveal unidirectional excess for this same channel. Indeed, Figure \ref{fig:diagnostics} shows the circular momentum arrows for both Max1 and Min1p2 samples, with the Max1 arrow measuring 0.308$^\circ$ (0.617R), while the Min1p2 arrow length is only 0.057$^\circ$ (0.115 R). 

Note that, for the 9 GeV energy bin, the KS tests for Min1 and Min2 samples show a strong anisotropic trend from the Min2 alone, with p-value reaching close to the 0.01 threshold. Only when considering the Min1 and Min2 sample together (i.e. Min1p2) does the significance of the anisotropic trend mount, suggesting the presence of anisotropic behavior, which the Rayleigh test was not able to resolve. Therefore, in this case, the data suggest significant deviation from the isotropic emission, which is not necessarily linked to a unidirectional excess.

\subsection{The Energy-dependent Anisotropic Trends}

The results from KS and Rayleigh test (Figure \ref{fig:ks-test} and \ref{fig:rayleigh-test}) highlight that the anisotropic trends are energy-dependent. This observation emphasizes the potential insights gained by considering the events' distribution within discrete energy bins, rather than integrating over the entire energy range provided by Fermi-LAT observations. The observed behavior could hint at a coupling between the solar magnetic field strength and the energy channels through which the $\gamma$-ray photons (produced via CR interactions with the solar surface) have a higher probability of escaping. Other factors, such as variations in the density and composition of the solar surface, or even fluctuations in the incoming CR flux, could also contribute to the observed anisotropic $\gamma$-ray emission. Further studies with larger event samples will be necessary to investigate those hypotheses. 

While the addition of data from yet another cycle observed with Fermi-LAT is likely to increase the number of measured events, the spatial concentration of those events may vary according to the solar cycle. It remains to be seen whether the Max2 period will produce a similar $\gamma$-ray emission pattern as observed during the Max1. 

The KS test showed more flexibility than the Rayleigh test to unveil deviations in the event distributions. However, the Rayleigh test provides valuable directional information (i.e. is able to resolve unidirectional excesses) as indicated by its p-value and the corresponding circular momentum arrow. When combined, the KS and Rayleigh tests constitute a powerful tool set to investigate and compare the strength of directional excess across different time windows and energy bins. Future studies can leverage these tests to further elucidate the mechanisms driving episodes of anisotropic $\gamma$-ray emission from the solar disk.

To better investigate the connection between solar activity, its magnetic configuration, and the observed $\gamma$-ray signatures, an advanced GeV $\gamma$-ray observatory with enhanced sensitivity will likely be required. Such an observatory would allow us to investigate solar emission with larger samples of $\gamma$-ray events and with improved reconstruction capabilities to better constrain event localization. This would facilitate the analysis and interpretation of the solar $\gamma$-ray activity, helping to refine our understanding and complement observations across different energy bands. \cite{GalCR-EOrlando2023} highlighted the significance of synchrotron radiation from galactic cosmic-ray electrons in the quiet Sun, suggesting that radio to X-ray observations can offer valuable insights into the interplay of CRs and solar magnetic fields. Furthermore, forthcoming X-ray and MeV-GeV observations from missions like FOXSI \citep{FOXSI-2014}, Gecco \citep{Gecco-2022}, ASTROGAM \citep{Astrogam-2017} and AMEGO \citep{AMEGO-2017}, will shed light on high-energy processes in the solar corona; while observations at TeV energies with HAWC \citep{HAWC-TeVSun-2018}, LHAASO \citep{LHAASO-2019}, SWGO \citep{SWGO-2019}, and ongoing MeV-GeV studies with Fermi LAT will further enhance our understanding of the solar disk emission at VHE.

\section{Time-Evolution of Solar Disk $\gamma$-ray Emission}
\label{sec:movie}

In this section, we expand on our visualization efforts and looked forward a temporal representations of the $\gamma$-ray data. Up to this point, our analysis has focused on subsamples aggregating $\gamma$-ray events over extensive periods, spanning multiple years (e.g., Max1, Min1p2). While this approach has been instrumental in identifying the existence of energy-dependent trends, it may obscure transient anisotropic patterns that emerge over shorter time periods. Such variations in $\gamma$-ray emissions could be diluted -or even completely masked- when data are summed over long time windows. To investigate the existence of transient patterns and gain deeper insights into the temporal dynamics of solar disk $\gamma$-ray emission, we now shift our focus to a time-resolved analysis. This approach will allow us to trace the evolution of anisotropic trends across the observational period from 2008 August to 2022 January. 

To provide a comprehensive view of the $\gamma$-ray distribution over time, we condense six different diagnostics of the solar disk activity (described in the following paragraphs) and follow their evolution along 2008 August to 2022 January. To account for the presence of energy-dependent features revealed in our previous analysis (sec. \ref{sec:diagnostics}), we have divided our sample into two categories: one for lower-energy events (\textbf{low-E}) and another for higher-energy events (\textbf{high-E}). For our time analysis, we incorporate photon counts within a specific time window, which is then shifted by 100 days per frame\footnote{We implement a 100-day sliding window for photon counts to balance between temporal resolution and the number of trials in our statistical analysis. For the low-E and high-E samples, \textit{n} amounts to 45 and 42, respectively.}, so that the statistical tests are evaluated over time.

\begin{figure*}
\begin{interactive}{animation}{movie-lowE.mp4}
\includegraphics[width=0.96\linewidth]{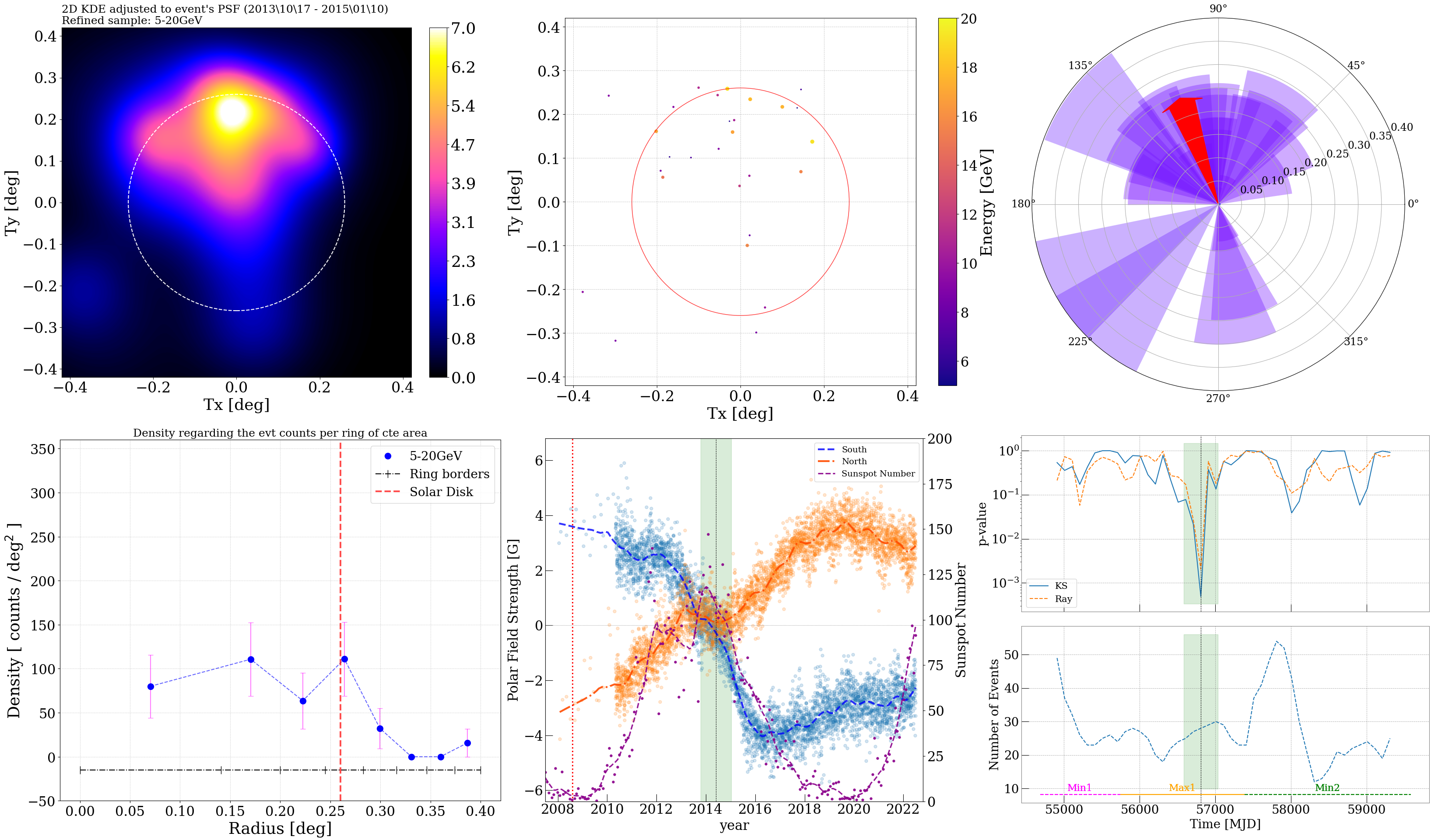}
\end{interactive}
\caption{This figure is available as an animation, and presents all diagnostic plots, from (i) to (vi), relating to the low-E sample, which include events with energies ranging from 5 to 20 GeV. In this particular frame, the diagnostic plots are derived from a time window centered on MJD 56807.6 integrating data across a span of 450 days, from 2013 Octobert 17 to 2015 January 10. This frame represents a period with the lowest p-values for the KS and Rayleigh tests, which coincided with the polar magnetic field flip at the maximum of cycle's 24 solar activity.}
\label{fig:movie-lowE}
\end{figure*}

\begin{figure*}
\begin{interactive}{animation}{movie-highE.mp4}
\includegraphics[width=0.96\linewidth]{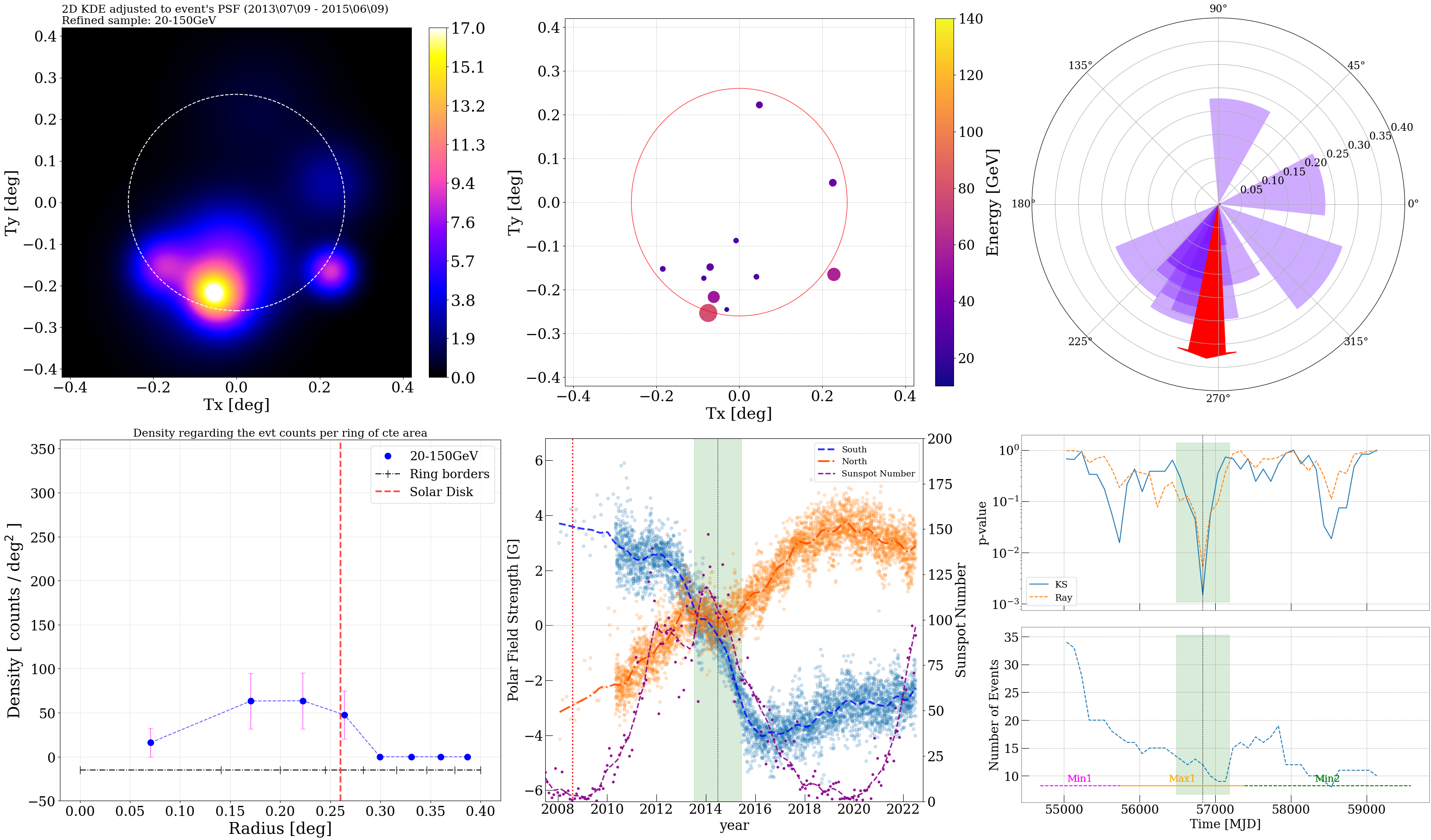}
\end{interactive}
\caption{This figure (available as an animation) presents all diagnostic plots, from (i) to (vi), relating to the high-energy (high-E) sample, which include events with energies ranging from 20 to 150 GeV. In this particular frame, the diagnostic plots are derived from a time window centered on MJD 56832.6, integrating data across a span of 700 days, from 2013 July 09 to 2015 June 09. This frame represents a period with the lowest p-values for the KS and Rayleigh tests, which coincided with the polar magnetic field flip at the maximum of cycle's 24 solar activity}
\label{fig:movie-highE}
\end{figure*}

For the low-E sample, we include events with energies ranging from 5 to 20 GeV. We compile these events over a 450 days time window and slide the window by 100 days for each subsequent frame (Figure \ref{fig:movie-lowE}). The high-E sample includes events with energies between 20 and 150 GeV. Given the lower frequency of these higher energy events and to ensure an adequate sample size, we use a longer time window of 700 days and slide it by 100 days for each frame (Figure \ref{fig:movie-highE}). The selection of 450 days and 700 days time windows for the low-E and high-E samples is motivated by achieving a balance between the number of available events and the intended time resolution for our analysis. 

For each frame we considered six types of analyses of both low-E and high-E samples: (i) a density plot, (ii) event localization on the solar disk, (iii) circular distribution of the $\gamma$-ray events, (iv) radial distribution of the events, (v) the polar magnetic field strength, and (vi) p-values from the KS and Rayleigh tests, along with the respective number of events used in these statistical tests.

In next items, we dive into the details of each plot featured in Figure \ref{fig:movie-lowE} and \ref{fig:movie-highE}. Essentially, those figures represent frames of an animation (provided as online material), produced when visualizing the sequence of slide frames played out\footnote{For visalization purpose, the animation available as online material considers frames slide by 20 days, to provide a smoother, more refined view of the time evolution.}. 

\begin{itemize}
    \item \textbf{(i) Density Plot:} the color code represents the density of events, smoothed using a 2D KDE function that accounts for individual event's PSF to compute the overall density distribution (see sec. \ref{sec:diagnostics}). This approach is meant to highlight regions with the highest concentration of events, while accounting for the event's uncertainty in position.

    \item \textbf{(ii) Localization of Events:} with each event's position plotted over the solar disk. The color and size of each event marker represent the event's energy, allowing a visualization of the relationship between event energy and spatial distribution. This comprehensive representation of the $\gamma$-ray events provides valuable insights into the correlation between energy and position, contributing to a more in-depth look into the anisotropic emission patterns.

    \item \textbf{(iii) Circular Distribution:} highlighting the $\gamma$-ray events' angular positions relative to the solar disk centre. The resulting vector (trigonometric momentum) measures the overall directional preference in the distribution of events. This vector visually represents the dominant direction or trend in the event positions, aiding in the interpretation of potential anisotropy in the $\gamma$-ray emission from the solar disk.
    
    \item \textbf{(iv) Radial Profile:} depicts the distribution of $\gamma$-ray events as a function of their radial distance from the center of the solar disk. This profile helps identify potential irregularities in the event distribution, as discussed in sec. \ref{sec:radial-dist}.

    \item \textbf{(v) Polar Magnetic Field Strength:} the north and south polar magnetic field strength (left-hand y-axis) are shown. To produce this representation, we rely on the Solar Dynamics Observatory (SDO) data from its Helioseismic and Magnetic Imager (HMI) instrument, which has been providing daily coverage since 2010. We use the \textit{drms}\footnote{The Data Record Management System (DRMS) developed by the Joint Science Operation Center (JSOC) at Stanford University.} Python package to access the HMI data \citep{sun-polar-Glogowski2019}. For the period from Fermi-LAT's launch in 2008 August to late 2010, we supplement the data using records from the National Solar Observatory (NSO), following \cite{sun-polar-Janardhan-2018}. The dashed and dotted-dashed lines represent south and north polar fields, respectively, averaged over approximately 20 days. We also show the monthly number of sunspots\footnote{ Sunspot counts reported at the Space Weather Prediction Center (SWPC): \url{https://www.swpc.noaa.gov/products/solar-cycle-progression}} (right-hand y-axis), tracking the occurrence of the cycle 24 solar maximum. A dashed dark-purple line represents the smoothed solar activity, which peaks around 2014 May-June. The green shaded region correspond to the time window for the current time frame.

    \item \textbf{(vi) P-values Time Window:} p-values derived from KS and Rayleigh tests are shown, with the number of events regarding each sample, all plotted against time (in MJD). Note that the event counts here are not corrected for exposure; hence, they do not represent fluxes. This count is intended to demonstrate the number of events used in our statistical test and support the derived conclusions. The green shaded region represents the time window used to select the $\gamma$-ray events for the corresponding plots. The low-E sample integrate events over a span of 450 days, while the high-E sample does so over 700 days. At the bottom of this plot, the solar activity cycles are marked as Min1, Max1, and Min2.  
\end{itemize}

With this approach we can resolve anisotropic episodes in time, consider energy dependencies, and explore possible connections with the solar cycles and magnetic configuration. The successful creation of these visualizations is only possible due to the well-reconstructed events in our refined sample (i.e., PSF better than 0.13$^\circ$) ensuring a meaningful representation of the events spatial-distribution along time. As a sound example of $\gamma$-ray time-analysis of the solar emission, \cite{TimLinden-2022} was the first to unveil the solar disk 1-10 GeV light curve over an entire solar cycle (see their Figure 5, where we contemplate the anticorrelation between the $\gamma$-ray flux and the solar activity) and also the first to investigate high-frequency variability of the solar disk $\gamma$-ray emission (i.e. looking for periodicity in flux variability) while tracking the potential origins for the signatures found in the power spectrum. In our case, by visualizing the evolution of $\gamma$-ray parameters associated to the solar disk (e.g. events distribution, density, radial profile of events, etc.), we look forward to valuable insides on the underlying nature of the observed emission.

\begin{figure}
    \centering
    \includegraphics[width=1.0\linewidth]{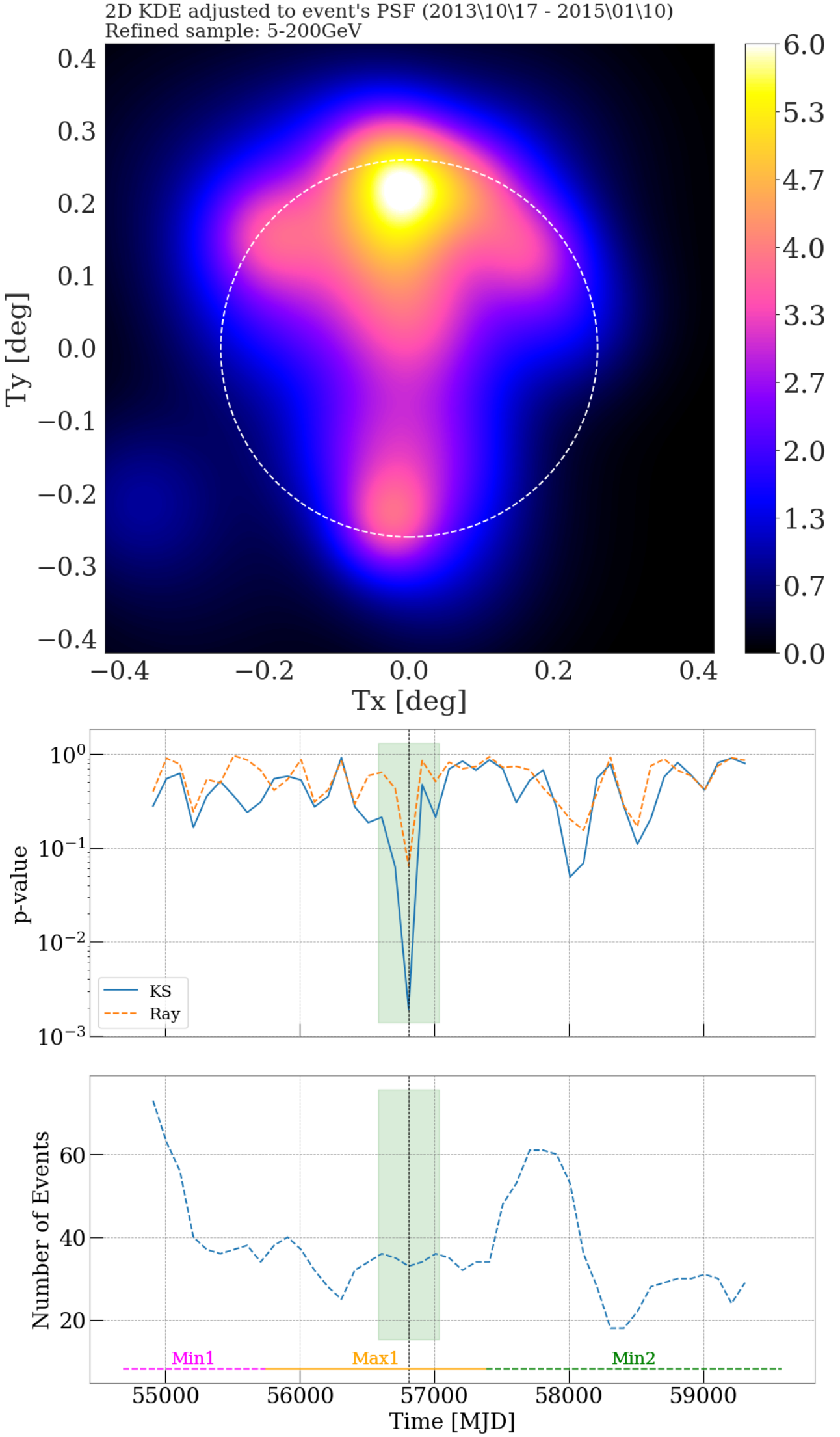}
    \caption{This figure presents diagnostic plots i) and vi) for the entire sample, which includes events with energies ranging from 5 to 150 GeV. In this case, the diagnostic plots are derived from a time window centered on MJD 56807.6, integrating data across a span of 450 days, from 2013 October 17 to 2015 January 10. This frame represents a period with the lowest p-values for the KS test, which coincided with the polar magnetic field flip at the maximum of cycle's 24 solar activity}
    \label{fig:movie-allE}
\end{figure}

In Figure \ref{fig:movie-lowE} and Figure \ref{fig:movie-highE} we highlight frames showcasing the most pronounced anisotropic trends in the low-E and high-E samples. From this perspective, we see that the lowest p-values occur at about MJD 56820 (2014 June) for both the KS and Rayleigh tests, and both the energy samples. 

With the time analysis, we now recognize an excess of 5-20 GeV events in the north pole, and a corresponding excess of events between 20 to 150 GeV at the south pole. In both energy samples, the KS and Rayleigh p-values go down to 10$^{-3}$ level, indicating significant anisotropic trends in the data. The Rayleigh test suggests unidirectional excess trends in the circular distributions, with trigonometric momentum pointing to opposite directions (north and south) for the low- and high-energy samples, therefore asymmetric regarding energy. This concentration of events in the solar poles is visible from the 2D KDE density plot, in agreement with the indications from the Rayleigh test. 

Notably, this energy-dependent, dipole-like pattern coincides with a period of sign-flipping in the dipole magnetic field  (refer to plot v) in Figs. \ref{fig:movie-lowE} and \ref{fig:movie-highE}). As pointed out by \cite{sun-polar-Janardhan-2018}, the northern polar field was nearly null from 2012 June to the end of 2014, indicative of unusual polar field behavior during this period. During the $\approx$13.3 years of Fermi-LAT observations (2008 August to 2022 January), the Sun went through a single episode of polar magnetic filed flip (i.e. a single solar maximum).

We can now investigate the following question: What is the likelihood that the strong anisotropic trends -observed in both the low-E and high-E samples coincide- with the solar maximum, only by chance?

In addressing this question, several factors need to be considered: \textbf{(1) P-value}. The p-value represents the likelihood of the Tx and Ty distributions being statistically similar (originating from from the same underlying population of events). For instance, we use the values derived from the KS tests, which are the most relevant/constraining; \textbf{(2) Number of Trials (n)}. To account for the number of trials, we will consider a  Bonferroni correction to adjust the p-values derived from the time analysis, as detailed in section \ref{sec:ks}. \textbf{(3) Coincidence with Solar Maximum}. Estimating the chance of these trends coinciding with the solar maximum is a crucial factor. Assuming a typical time span for the solar maximum (or for the polar magnetic field-flip) to be approximately one year, we roughly estimate a 1/13 chance of observing a significant anisotropic trend during this period, purely by chance (\textit{p-smax}). \textbf{(4) Independence of Low-E and High-E Results}. In principle, there is no known relation between these energy ranges. Indeed, we want to evaluate if that is the case, in connection with the solar maximum (polar field sign-flip). \textbf{(5) Simultaneity of Trends}. Notably, these anisotropic trends, drawn from independent samples, occur nearly simultaneously, exhibiting energy-dependent asymmetry.

Considering these factors, the chance probability (\textit{p-lowE}) of observing the low-E anisotropic trend during the solar maximum can be calculated as the product of the p-value, the number of trials (n), and the chance of coincidence with the solar maximum (\textit{p-smax}); This calculation results in \(0.00049 \times 45 \times 1/13 \approx 0.0017\). Similarly, for the high-E sample (\textit{p-highE}), the calculation yields \(0.0015 \times 45 \times 1/13 \approx 0.0051\).

Taking into account the independence of the low-E and high-E results and their simultaneous occurrence, the combined probability of observing strong anisotropic and asymmetric trends in the solar $\gamma$-ray emission, coinciding with the solar maximum, is estimated as \((\textit{p-lowE} \times \textit{p-highE}) = 8.7 \times 10^{-6}\). This result, carefully accounting for the trial factor, corresponds to a statistically significant \(3.7\sigma\) signature of asymmetric $\gamma$-ray emission during the solar maximum.

\textbf{Time Analysis for the Entire Sample:} In Figure \ref{fig:movie-allE}, we highlight the most pronounced anisotropic trends observed when considering the entire sample, including events with energies ranging from 5 GeV to 150 GeV (i.e., both the low-E and high-E samples combined). In this case, the lowest p-value occurs at about MJD 56812 (2014 June) for the KS test. Note, the Rayleigh test does not show signs of unidirectional excess, as expected in case of bipolar distribution of events; Only when splitting between low-E and high-E events the Rayleigh is able to unveil the unidirectional excess to the north and south, respectively. Therefore, we note that the KS test demonstrated reliable sensitivity to unveil the presence of bipolar emission patterns aligned to the rotation axis. In this test, the anisotropic signature revealed by the KS test does not depend on the energy sampling and reinforces the robustness of our detection.  

For a more detailed visual inspection of the events distribution evolving over time, we refer the reader to the corresponding animations in the online supplementary material, associated to Figures \ref{fig:movie-lowE} and\ref{fig:movie-highE}. 

\section{Conclusions}

In this study, we investigated the properties of the solar disk $\gamma$-ray emission, with data obtained from the Fermi-LAT satellite over 2008 August to 2022 January. Our analysis shows robust evidence indicating anisotropic and energy-dependent asymmetries, offering valuable insights into its nature and extending \cite{ElenaOrlando-Sun-EGRET-2008,KennyNg-GammaTvarSun-2016,TimLinden-2018,TimLinden-2022} pioneering works.

To ensure the reliability of our $\gamma$-ray sample, we have applied a series of comprehensive data cuts that systematically exclude times when the solar path is linked to known and potential contaminants. This includes the solar transit through the galactic-disk background and close encounters with the Moon. It also includes coincidences with GRBs, transits associated with known $\gamma$-ray sources (4FGL-DR3, 2FAV, 1FLT, and 2BIGB), as well as transits regarding blazars and blazar candidates in general (even for those not yet detected in $\gamma$-rays). 

The initial phase of our analysis focuses on a careful examination of the GeV $\gamma$-ray distribution over the solar disk, evaluating its relation to energy and solar cycles. To that end, we consider $\gamma$-ray events within the following solar activity cycles: Min1, Max1, and Min2. We combine both KS and Rayleigh tests to investigate the existence of anisotropic trends and energy dependencies within each cycle. With this approach, we found relevant anisotropic trends in $\gamma$-ray emissions, most notably at the Max1 50 GeV channel, with excess counts toward the solar south pole. We also recover a significant anisotropic trend at the 9 GeV energy bin for the Min1p2 sample (a combination of Min1 and Min2 samples), revealing compelling evidence of anisotropic and energy-dependent trends in the solar disk $\gamma$-ray emission. Through this initial analysis, we demonstrate that a combination of the KS and Rayleigh tests provides a powerful tool to investigate the solar disk emission across different time windows and energy bins.

In the subsequent phase of our study, we divided our sample into lower (5-20 GeV) and higher (20-150 GeV) energy events. Our aim was to investigate energy-dependent features over time, which we achieved by integrating the $\gamma$-ray events over a moving time window and storing the results as frames. This strategy enabled us to build a dynamic visualization of the $\gamma$-ray distribution over time, essentially an animation of our data exploration. This approach yielded a unique perspective on the temporal evolution of the observed anisotropies.

Through our visualization we were able to pinpoint the most prominent anisotropic signatures in both low- and high-energy samples, which emerged around MJD 56820 (approximately 2014 June). 

This period shows an intriguing dipole-like pattern, which is energy dependent and coincident with the polar magnetic field flip during the cycle 24 solar maximum. This observation reinforces the idea of a potential link between $\gamma$-ray emission, the solar magnetic configuration, and solar cycles. Our analysis estimates a 3.7$\sigma$ association between the asymmetric $\gamma$-ray emission and the polar magnetic field flip at the cycle 24 solar maximum. Looking ahead, we anticipate that the Fermi-LAT coverage of the ongoing Max2 cycle (from January 2022 onwards) will clarify whether the dipole pattern is typical of the solar maximum.  

Our study presents a step forward in understanding solar $\gamma$-ray emissions and introduces new questions in the emerging field of Very High Energy Solar Astrophysics. We are likely observing the complex interplay of solar magnetic fields, solar atmospheric conditions, and incoming CRs, which includes fluctuations in CR flux and composition. There is a vast scientific potential to exploring the physical processes that shape and influence the energy channels through which $\gamma$-rays are more likely to escape the solar surface. Based on our findings, we envisage that future $\gamma$-ray missions could benefit from the ability to provide real-time data on solar $\gamma$-ray activity, serving as extra ingredient for solar forecasting models.

\begin{acknowledgments}
We thank ApJ's anonymous Referee for the careful review and insightful comments. We acknowledge Melissa Pesce-Rollins, Philippe Bruel, Matthew Kerr, and especially Francesco Loparco for useful comments at the late stage of the manuscript. BA thanks the University of Trieste and the INFN - IT for the ``Assegno di Ricerca" grant (2021 July to 2022 June) that supported the initial stages of this research, under the ASI-INAF-CUP F82F17000240005 research project ``Gamma rays from the Quiet Sun and Diffuse Emissions." Further thanks go to the Institute of Astrophysics and Space Sciences (IA), University of Lisbon. B.A. is currently a Marie Sk\l{}odowska-Curie Postdoctoral Fellow at IA, funded by the European Union's Horizon 2020 research and innovation program under the MSCA agreement No. 101066981. BA thanks the support from `Fundação para a Ciência e a Tecnologia' (FCT) through the research grant UIDP/04434/2020 (DOI: 10.54499/UIDP/04434/2020). E.O. acknowledges the NASA grant No. 80NSSC20K1558. We acknowledge the \textit{Centro de Computa\c{c}\~{a}o John David Rogers} (CCJDR) at IFGW Unicamp, Campinas-Brazil, for providing access to the Feynman and Planck HPC Clusters, essential for the parallel computation of $\gamma$-ray light curves using the Fermi Science Tools. Additionally, part of this work utilizes archival data, software, or online services from the Space Science Data Center - ASI. The \textit{Fermi} LAT Collaboration acknowledges generous ongoing support from a number of agencies and institutes that have supported both the development and the operation of the LAT and scientific data analysis. These include the National Aeronautics and Space Administration and the Department of Energy in the United States, the Commissariat \`a l'Energie Atomique and the Centre National de la Recherche Scientifique / Institut National de Physique Nucl\'eaire et de Physique des Particules in France, the Agenzia Spaziale Italiana and the Istituto Nazionale di Fisica Nucleare in Italy, the Ministry of Education, Culture, Sports, Science and Technology (MEXT), High Energy Accelerator Research Organization (KEK) and Japan Aerospace Exploration Agency (JAXA) in Japan, and the K.~A.~Wallenberg Foundation, the Swedish Research Council and the Swedish National Space Board in Sweden. Additional support for science  analysis during the operations phase is gratefully acknowledged from the Istituto Nazionale di Astrofisica in Italy and the Centre National d'\'Etudes Spatiales in France. This work performed in part under DOE Contract DE-AC02-76SF00515.
\end{acknowledgments}

%

\vspace{5mm}
\facilities{Fermi-LAT, HMI, SDO}


\software{astropy \citep{astropy:2013, astropy:2018, astropy:2022},  
          sunpy   \citep{sunpy_community2020, SunPy-Mumford2020},
          scikit-learn \citep{scikit-learn},
          scipy  \citep{SciPy-2020}
          fermi-sciencetools \citep{FermiTools:2019}
          }

\bibliography{bibliograph}{}
\bibliographystyle{aasjournal}



\end{document}